\documentclass{aa}  

\usepackage{graphicx}
\usepackage{txfonts}
\usepackage[]{natbib} 
\begin{document}

   \title{Radio emission during the formation of stellar clusters  in M33}

   \author{Edvige Corbelli
          \inst{1}
          \and
            Jonathan Braine 
           \inst{2}
           \and
           Fatemeh S. Tabatabaei
                    \inst{3,4,5}
          }

   \institute{
             INAF-Osservatorio Astrofisico di Arcetri, Largo E. Fermi, 5,
             50125 Firenze, Italy\\
             \email{edvige@arcetri.astro.it} 
        \and
             Laboratoire d'Astrophysique de Bordeaux, Univ. Bordeaux, CNRS, B18N, 
                   all\'ee Geoffroy Saint-Hilaire, 33615 Pessac, France\\
          \and
                {School of Astronomy, Institute for Research in Fundamental Sciences, P.O. Box 19395-5531, Tehran, Iran}\\
               \and
                     {Instituto de Astrof\'isica de Canarias, 38205 San Cristóbal de La Laguna, Tenerife, Spain}\\
                     \and
                          {Max-Planck-Institut f\"ur Astronomie, K\"onigstuhl 17, 69117 Heidelberg, Germany}\\                  
}

   \date{Received ....; accepted ......}

 
  \abstract
  { }
   {We investigate  thermal and non-thermal radio emission associated with the early formation and evolution phases  of 
   Young Stellar Clusters (YSCs) selected  by their MIR emission at 24~$\mu$m in M33. We consider regions in their early formation period, 
   compact and totally embedded in the molecular cloud, and in the more evolved and exposed phase. 
    }
   {Thanks to recent radio continuum surveys  between 1.4 and 6.3~GHz 
   we are able to find radio source counterparts to more than 300 star forming regions of M33. We identify the thermal free-free component 
  for YSCs and their associated molecular complexes using the H$\alpha$ line emission.  
  }
   {A cross-correlation of MIR and radio continuum  is established from bright to very faint sources, with the MIR-to-radio emission ratio that 
   shows a slow radial decline throughout the M33 disk. We proof the nature of candidate embedded sources by recovering the associated faint radio continuum 
   luminosities.  By selecting exposed YSCs with reliable H$\alpha$ flux, we establish and discuss the tight relation between H$\alpha$ and the total radio 
   continuum at 5~GHz over 4 orders of magnitude.  This holds for  individual YSCs as well as for the  molecular clouds hosting YSCs and allowed us to calibrate
   the radio continuum  -  star formation rate  relation at small scales.
    On average about half of radio emission at 5~GHz in YSCs is non-thermal with a large scatter. For exposed but  compact YSCs  and their molecular clouds
    the non-thermal radio continuum fraction  increases with source brightness,  while for  large HII regions the non-thermal fraction is lower and shows no 
    clear trend. This has been found  for YSCs with and without  identified SNRs and underlines the possible role of massive stars in triggering particle 
    acceleration through winds and shocks: these  particles  diffuse throughout the native molecular cloud prior to cloud dispersal.
   }
   {}


\maketitle
%

\section{Introduction}

The radio continuum-infrared (hereafter IR) correlation in galaxies has been investigated since the data release by IRAS mission \citep[e.g.][]{1985ApJ...298L...7H,
2003ApJ...586..794B}.
In the past 35 years a lot of effort has been spent to investigate the frequency ranges in the IR and radio continuum for which the correlation
is the tightest and how this depends on spatial scale. In the IR domain the focus has been to distinguish warm dust emitting in the mid-infrared (MIR)
and heated by star formation, from cold dust heated by the interstellar radiation field (ISRF) and emitting in the far-infrared (FIR). 
In the radio continuum  both thermal (free-free) and non-thermal (synchrotron) emission are  linked to the formation of stars since this process is responsible for 
the free electrons, as well as for turbulent magnetic field amplification and cosmic ray (hereafter CR) production
 \citep[e.g.][]{2013A&A...556A.142S,2016ApJ...827..109S,2013A&A...552A..19T}.   
A key question is  the spatial scale involved  in the dust-to-radio correlation and to this purpose nearby galaxies such as LMC, M33, M51 etc.. have been 
analyzed by using wavelet decomposition and   maps at several spatial resolutions,  from 50~pc out to the kpc scale, for the radio and IR emission
\citep{2006MNRAS.370..363H,2007A&A...466..509T,2011AJ....141...41D,2013A&A...557A.129T}. On large scales the ordered 
magnetic field seems to plays a major role in driving non-thermal emission, especially where cosmic rays are able to diffuse 
away from star forming sites before losing their energy. Thus the magnetic field might be coupled to cold dust by density enhancement in the diffuse 
ISM \citep{2017MNRAS.471..337B}. In M51 \citet{2011AJ....141...41D}  explored how
the relation between the MIR and radio continuum depends on the local environment such as spiral arms, interarm regions, outer versus inner disk and found that
a linear relation  holds only along the spiral arms with radio synchrotron emission being suppressed in the central regions.

On small scales  it is interesting to study the  non-thermal and thermal component and how these relates to
the warm dust  as a function of  the  ISM properties and environment in the galactic disk. 
Data in the radio continuum at 1.4~GHz and at 60~$\mu$m for  the LMC at spatial resolution between 50~pc and 1.5~kpc has been analyzed by 
\citet{2006MNRAS.370..363H}. They found that on small scales the thermal radio emission correlates better with the warm dust component while the non-thermal
FIR correlation breaks down below a certain scale possibly because of the large diffusion length of cosmic rays. 
More recently, a work on IC10 by \citet{2017MNRAS.471..337B} underlines that also at smaller scales, between 50 and 200~pc, the non-thermal radio emission
establishes a better correlation with the  dust emission at 70~$\mu$m than with the warm dust at 24~$\mu$m.  However there are no studies analyzing
individual star forming regions, where the 24~$\mu$m emission peaks.

The large scale magnetic field in M33 exhibit a  spiral structure with a  decrease in the radio thermal fraction   
going  radially outwards \citep{2007A&A...472..785T}. At kpc scales in M33  the warm dust-thermal radio correlation is stronger than the cold dust-non 
thermal radio correlation  \citep{2007A&A...466..509T}, although a cold dust  non-thermal correlation still holds.  
Analysis of the radio emission in the closest spiral galaxies \citep{2013A&A...557A.129T} has show that the non thermal-IR correlation is weaker in M33 than in 
M31 on large scales, but it is stronger than in M 31 on scales $<$1 kpc. This has been explained by the smaller propagation length of CR electrons in M33, 
due to its turbulent magnetic field structure.  At smaller scales the turbulent magnetic field  might become more important 
as turbulent gas motion injects energy into the ISM \citep{2008A&A...490.1005T}.  If this is a general property  of the M33's ISM, the non thermal--IR correlation 
should also hold on scales where the turbulent magnetic field is strongest, i.e., in the star forming clouds. 

For nearby galaxies the sensitivity and spatial resolution of infrared and radio surveys is now  sufficiently high that it is possible to isolate  individual star forming regions   
with stellar masses as low as a few hundreds solar masses \citep{2011A&A...534A..96S,2019ApJS..241...37W}. This means that instead of focusing on the
spatial scale and on a pixel by pixel analysis one can examine the radio and IR emission in individual star forming regions and how this depends on characteristic 
properties of  the  young  cluster such as mass  and age and more in general on  the energy input that  massive stars can provide. 
In this paper we analyze the radio continuum and MIR emission in individual star forming sites of M33, 
the closest blue spiral galaxy. This will allow us to test whether  the correlation 
between the non-thermal radio emission and the MIR emission found down to 200~pc scale \citep{2013A&A...557A.129T} holds also at smaller scales
between the emission of individual young stellar clusters.
Previous works have presented a catalogue of infrared selected star forming regions
across the whole disk of M33 \citep{2011A&A...534A..96S} which has been complemented by a catalogue of GMCs \citep{2017A&A...601A.146C} and recently 
a list of radio-continuum sources in the M33 area became available \citep{2019ApJS..241...37W}, some of which might be related to the M33 disk. 
The spatial resolution at 24~$\mu$m is  comparable to that of the recent radio-continuum surveys.  Our aim is to investigate the correlations between the
radio continuum and other star formation tracers such as H$\alpha$ or  MIR emission and investigate the dominant mechanism that provides radio emission, if thermal or
non thermal, across a variety of star forming regions. While it is clear that the turbulent magnetic field is enhanced in 
star forming regions  it is less clear how this depends on the characteristic of the region and its location in the disk.
Furthermore the creation and propagation of  CRs might depend on the galaxy \citep{2013A&A...557A.129T} and on characteristics of the SF region.
 
Currently most of the models involving  CRs acceleration are based on supernova (SN) explosions and their remnants (SNRs).  In 
particular these models can explain  the CR production rate, the composition and anisotropy in our Galaxy. However, the steepness and the presence of 
breaks in SN spectra as well as the difficulties  in explaining the very energetic CRs opens the possibility to other types of acceleration mechanisms. 
Moreover  the duration of the SN shock might be too short to explain the propagation away from the remnants that is needed to explain some observation.
Over the last decade, space- and ground-based telescopes have revealed many classes of Galactic $\gamma$-ray sources, such as clusters of young stars,
which might be associated to CR accelerators. As underlined recently by \citet{2019NatAs...3..561A}  the acceleration could take place in the vicinity of the stars  
or in superbubbles,  caused by interacting winds of 
massive stars as those in OB associations. A fraction of stellar wind mechanical energy may be transferred  to relativistic CRs by diffuse shock acceleration 
at the wind boundary \citep{1983SSRv...36..173C,2004A&A...424..747P,2019A&A...630A..72P}.
These will spiral around the fluctuating amplified magnetic fields  of the star forming region emitting non-thermal radio waves.  To test these possibilities recently  identified  
SNRs in star forming regions of M33  \citep{2019ApJS..241...37W} can be used for assessing their role in enhancing the non thermal radio emission in SF regions .

The plan of the paper is the following: In Section~2 we present the data and the method to establish the association between  MIR and radio sources. In Section~3
we present the correlation between radio continuum and  MIR  fluxes in exposed YSCs  and in the embedded star forming sites in M33. In Section~4
we derive the non-thermal fractions and the relative implications. In Section~5 we present some preliminary results on radio emission from molecular clouds,  and in
 Section~6 we discuss the link between the star formation  and  radio emission. Section 7 summarizes the main results of our analysis.

\section{The YSC sample and  multiwavelength data }

In this Section we describe how we select candidate YSCs of M33 during their formation and early evolutionary phases  and  how we retrieve the  
emission at other wavelengths. We use three catalogues now available for M33: a 24~$\mu$m source catalogue \citep{2011A&A...534A..96S}, 
a giant molecular cloud (GMC) catalogue \citep{2017A&A...601A.146C} and a radio continuum source catalogue \citep{2019ApJS..241...37W}.
The association is done by matching the positions in the sky of  the objects in these catalogues because the shape and extend of the emission
of a star forming region in the MIR and in the radio continuum  might not be the same. The hot dust for example might be located
closer to the GMC center or to the shell of an expanding ionized bubble while non-thermal radio continuum emission might peak close to a SNR.  

\subsection{The  MIR source catalogue}

Dust emission at MIR wavelengths has been investigated through
data of the  Multiband Imaging Photometer (MIPS) on board {\it Spitzer} processed as
described  by \citet{2007A&A...476.1161V}. 
\citet{2017A&A...601A.146C} selected 630 MIR sources from the list of \citet{2011A&A...534A..96S} that are candidate YSCs   
in the early formation and evolutionary phases: the results of
the classification, together with  the most relevant parameters of the GMCs and YSCs,  can be found in their online tables.  
The YSCs were sorted into classes based on their emission in the MIR, FUV, and H$\alpha$ bands and according to the association with GMCs.
The classification helps in drawing a possible evolutionary sequence and the relative timescales see \citep{2017A&A...601A.146C} for more details.
We recall the classification scheme of the MIR sources and the number of MIR sources in each category:

\begin {itemize}
{\item class b}:  97  MIR sources, associated with GMCs, with no optical counterpart (with no or very weak H$\alpha$ peak)    
{\item class c}:  410  YSCs with optical  counterpart  
\begin {itemize}
{\item c1}:  55 YSCs associated with GMCs with coincident H$\alpha$ and MIR emission peaks but FUV emission  spatially shifted or absent 
{\item c2}:  218 YSCs associated with GMCs with coincident H$\alpha$, FUV and MIR emission peaks  
{\item c3}:  139 YSCs not associated with GMCs but with optical and FUV counterparts; these often have weak H$\alpha$ emission 
\end {itemize}
{\item class d}:  19 MIR sources associated with GMCs which are ambiguous for b or c1/c2 class
{\item class e}:  104 MIR sources not associated with GMCs and with no H$\alpha$ emission, some FUV may be present 
\end {itemize}
 
The b-c1-c2-d-type are sources associated with giant molecular clouds (GMCs). The
b-type are called embedded because there is no FUV or H$\alpha$ counterparts, while c-type are exposed star forming regions with FUV and H$\alpha$ 
counterpart and some of them (c1 and c2) are associated with GMC while others (c3) are not. The d-type are ambiguous between b-and c-type. 
From now on we shall refer to the c-class as exposed YSCs and to the b-type as embedded YSCs. Extinction is  high for b-type sources with weak 
or no UV/optical counterpart, which likely represent the early phases of SF. The c1-type YSCs have visible H$\alpha$ but not FUV emission and show on 
average higher extinction than the c2-type YSCs, where  FUV emission is also detected. The c1-type YSCs may represent YSCs at an earlier stage than c2-type  
YSCs, even though the YSC age determination is not
precise enough to separate  these two classes. The most luminous MIR sources are of c2-type and are clusters that have very likely completed the formation process.
This is also the largest class of MIR sources for which the coincident peaks in the UV and H$\alpha$ bands enabled more precise estimates of the age
and mass of the associated YSC.  

The e-type sources have only MIR emission but unlike the b-type these do not have any associated GMC. 
We marginally consider  the e-type in this paper. As shown in subsequent Sections and also in the paper by \citet{2019A&A...622A.171C}
where deep CO searches have been carried out, the e-type are highly contaminated by background sources and thus most of them do not
belong to M33. Background contamination is particularly relevant for dim sources, with flux at 24$\mu$m smaller than 5~mJy, and no associated molecular cloud or
H$\alpha$ emission such as type-e sources  \citet{2019A&A...622A.171C}. We cannot exclude, however that a few of them might be embedded YSCs in
low mass molecular clouds. 

\subsection{The molecular cloud catalogue}

Using the IRAM 30m CO J=2-1 data cube of M33, we identified 566 GMCs, as described in \citet{2017A&A...601A.146C}, and whose properties
are listed in the GMCs catalogue.
The GMC masses are  between 2$\times 10^4$ and 2$\times 10^6$ M$_\odot$ and GMC radii between 10 and 100~pc The number of GMCs 
above  the survey completeness limit (M$_{H_2}\ge 6.3\times 10^4$~M$_\odot$) is 490. \citet{2017A&A...601A.146C} classified the GMCs as 
non-starforming (class A), with embedded SF (B), or with exposed SF (C). A few ambiguous cases are in class D.
The FUV, H$\alpha$ and MIR emission maps of M33 have been used to classify the GMCs. 

The majority of the  MIR sources  in the inner zones (R$<$4~kpc)
lie  within a GMC boundary while this is not true for MIR sources in the outer disk.
There is an extraordinary spatial correspondence between the GMCs and the distribution of atomic hydrogen 
overdensities in the inner zones.
In the outer zone there are fewer GMCs, possibly because of a steepening of the molecular cloud mass spectrum, with a larger fraction of clouds
being below the survey completeness limit.

\subsection{The radio source catalogue}

Recently using  JVLA radio continuum maps of   M33   at 1.4 and 5~GHz    
 at   a spatial resolution of about 6~arcsec,  \citet{2019ApJS..241...37W}   catalogued 2875 sources with fluxes between  2 and 33000~$\mu$Jy (except for one
with a flux of  about 0.17~Jy) with a completeness limit around  300~$\mu$Jy. Most of the luminous radio sources are stellar-like objects, background QSOs
or MW stars; a few  are sources associated with the two most prominent HII regions of M33, NGC 604 and NGC 595, which we exclude from the current analysis because of their
extended and complex structure.  \citet{2019ApJS..241...37W} have also identified SNR using the optical catalogues of 
\citet{2010ApJS..187..495L,2014ApJ...793..134L}. Background contamination is  present at all flux densities in radio but the number of background sources dominates
over local sources in number as the flux gets below  300~$\mu$Jy. To identify M33 sources it is therefore mandatory to use identification of other counterparts
with line emission such as H$\alpha$ and molecular clouds.

\subsection{Emission at other wavelengths }

For MIR sources we have recovered the emission at other wavelengths, 
such as H$\alpha$,   by aperture photometry which has been described in detail by \citet{2011A&A...534A..96S}. We noticed that the for radio sources coincident with MIR sources
the H$\alpha$ emission in \citet{2019ApJS..241...37W} catalogue was much lower that that associated to MIR sources  by \citet{2011A&A...534A..96S}, especially for bright sources.
Accurate H$\alpha$ photometry is   needed to estimate the thermal fraction of radio emission  from YSCs.
Therefore to trace the ionized gas  associated to radio sources we perform new aperture photometry,  centering 
a circular aperture on radio source coordinates. We set the aperture  radius equal to 1.5 times the mean between the semiminor and semimajor source axis, and subtracting 
the local background  \citep[see][for more details on aperture photometry]{2011A&A...534A..96S}.
As described by \citet{2000ApJ...541..597H} M33 has, in fact, a non negligible diffuse fraction of ionized gas due to leakage of ionizing photons from HII regions and 
to massive stars in the field. An aperture larger than the source radius account for  the position of the source center and extension which 
may change with wavelength. We adopt the narrow-line H$\alpha$ image of M33 obtained by \citet{1998PhDT........16G} and described in detail in \citet{2000ApJ...541..597H}.  
As suggested by the authors a 5$\%$ contamination by NII has been considered  and accounted for.     

Additional radio observations of M33 were carried out with the Karl Jansky Very Large Array (JVLA) 3 at C-band (5.5-7.5
GHz) in D configuration as part of the Cloud-scale Radio Survey of Star Formation and Feedback In Triangulum galaxy (CRASSFIT, Tabatabaei et al. in prep.) 
between November 2011 and March 2012. Two continuous base-bands (of 1024 MHz) were tuned at 5.5-6.5 GHz and 6.5-7.5 GHz, each with 8 sub-bands of 
128 MHz. A mosaic of 5x5 was used to cover the inner $18' \times 18'$ star forming region including the main spiral arms with similar sensitivity. 
The total on-source observing time per pointing was about one hour leading to an rms noise of 6 $\mu$J per $9".35$ beam in the final image.
The two sources 3C 48 and 3C 138 were used as the primary flux density and the antenna gain phase calibrator during the observations. 
The data were reduced in the Obit package (Cotton, 2008, PASP, 120, 439) using the standard VLA calibration procedure. 
Each pointing was imaged and then combined into a mosaic using the wideband Obit imager MFImage (Cotton et al. 2018, ApJ, 856, 67).
We shall refer to this dataset as the 6.3~GHz map. The negative sidelobes sometime visible in the White et al.  maps are not present in the 6.3~GHz
map which can be used for deeper and more sensitive searches of YSCs radio counterparts. 

In performing aperture photometry for radio sources on the 6.3~GHz map we use circular apertures with aperture radius equal to the  source radius at 1.4~GHz convolved 
with the beam of the 6.3~GHz map.  We subtract the local background, having corrected  the 6.3~GHz map  for missing short spacing, 
using the 'mode' function  in a  3 pixel wide annulus at a distance of at least 4 pixels from source aperture boundary.

We would like to underline here that we have also performed aperture photometry without background subtraction both on H$\alpha$ and on radio continuum maps.
The results discussed in the next Sections remain unchanged since flux variations are marginal with the largest flux increase being that relative to the 6.3~GHz data.
In Section~4, where we separate the thermal and non-thermal  radio flux at 5~GHz using H$\alpha$ emission, we also show results  for a varying aperture size and   
using a smoothed version of the H$\alpha$ image. This has been obtained by  convolving the original map (at about 3~arcsec spatial resolution) with a gaussian function of  
6~arcsec FWHM, which is the spatial resolution of the 24~$\mu$m and 5~GHz  maps.

Photometry for emission within GMCs is carried out using the the irregular GMC shape recovered by \citet{2017A&A...601A.146C}. The shapes and GMC images are available  
for a subsample of them in the on-line version of \citet{2012A&A...542A.108G}. No background is subtracted in this case although some low level of
diffuse emission might be associated with the atomic gas in the M33 disk along the  line of sigh to the GMC.

\subsection{Uncertainties}

The 1.4~GHz map of \citet{2019ApJS..241...37W} is less affected by sidelobes than the 5~GHz map and we trust the sources identified in the catalogue at 1.4~GHz.
 The source flux at 5~GHz  can be recovered using the catalogued spectral indexes and flux at 1.4~GHz. This is in good agreement with the flux recovered at 6.3~GHz
 if we use the 6.3~GHz map before corrections for short spacing are applied.  
 As noticed by \citet{2019ApJS..241...37W}, source fluxes at 1.4~GHz are lower than those of \citet{1999ApJS..120..247G}  and to get an agreement
a positive correction is needed. Increasing the 1.4~GHz fluxes of  \citet{2019ApJS..241...37W} by about 30$\%$  brings them 
in agreement  with the  fluxes recovered using the 1.4~GHz map of \citet{2007A&A...472..785T} and the unpublished JVLA map at the same frequency 
(Tabatabaei et al. in preparation). Moreover, in the radio catalogue the HII region spectral indices are mostly positive, with a mean value of about 0.1, 
while we expect them to be  negative  because  optically thick radio emission on the scale of tens of parsecs is unlikely. 
 Increasing the 1.4~GHz fluxes   by 30$\%$  implies  spectral indices lower by about 0.2 and hence more physically reasonable. However, in this paper
 we use the 1.4~GHz data without applying any correction but
we shall focus more on the high frequency emission at 5 and 6.3~GHz, for which no calibration corrections are needed.
 We use the \citet{2019ApJS..241...37W} data considering  30$\%$ calibration errors  in addition to photometric and spectral index uncertainties.

Photometric  errors for H$\alpha$ are negligible and calibration errors are of order 5$\%$. 

\section{Radio counterparts to MIR sources}

\begin{figure}
\hspace{-0.5 cm}
\vspace{2 cm}
\includegraphics[width=10. cm]{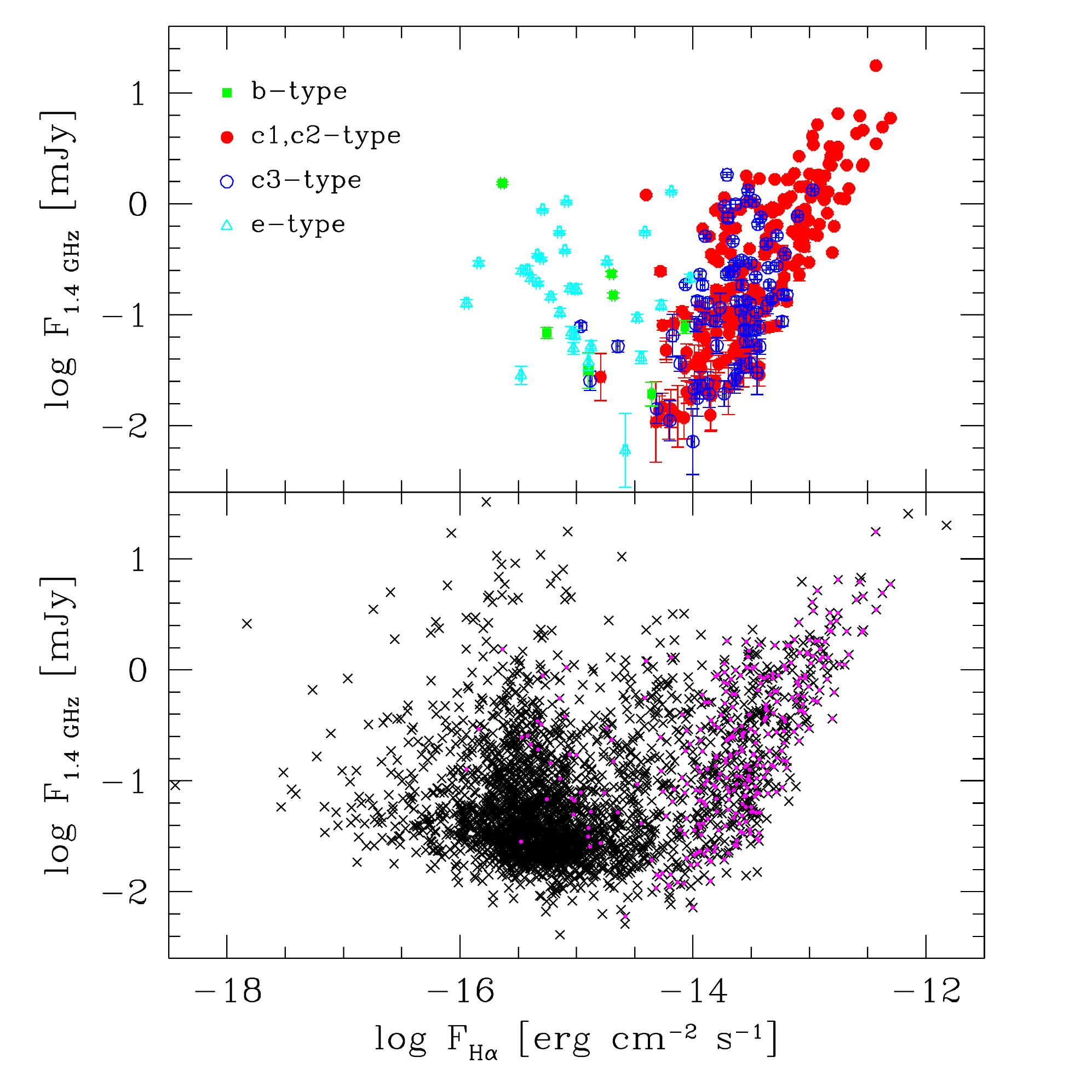}
 \caption{ The bottom panel shows the radio flux at 1.4~GHz of sources in the White et al. catalogue versus the associated H$\alpha$ flux in the same catalogue. The magenta color
 highlights sources in the radio catalogue which are associated with MIR sources in the catalogue of \citet{2011A&A...534A..96S}. These are plotted also in the upper panel where  
  filled (green and red) and open (blue and cyan) symbols are used for  MIR sources  associated  and not-associated with known GMCs respectively. The color coding,  as indicated  
  in the upper left corner,  outlines the classification scheme of \citet{2017A&A...601A.146C} (see text for details),  with blue and red colors marking
  MIR sources with a clear counterparts in H$\alpha$ (c-type).
  }
\label{RHa_all}
\end{figure}

To have an overview of the radio sources in the M33 area we plot
in Figure~\ref{RHa_all}   the radio flux at 1.4~GHz for  all radio-sources in the  \citet{2019ApJS..241...37W} 
catalogue versus H$\alpha$ emission  given in the same catalogue.  
There are clearly two distinct source populations, one for which   the radio and H$\alpha$ flux establish a correlation and the other, 
more numerous and with  lower H$\alpha$ fluxes, for which the correlation
is  lost.  Because of noise and diffuse emission H$\alpha$ fluxes  below 10$^{-15}$~erg~s$^{-1}$~cm$^{-2}$ are not peaks in the map. At the 
distance of M33 (840~kpc) this implies that we can only measure luminosities above  10$^{35}$~erg~s$^{-1}$ equivalent to  a source as faint as a B1-type star. 
The value of 10$^{-15}$~erg~s$^{-1}$~cm$^{-2}$ is also the 3-$\sigma$ limit for H$\alpha$ reliable counterparts of MIR 
selected sources using the  H$\alpha$ image we adopt   \citep{1998PhDT........16G} . Only in less crowded regions far from the 
center, where the diffuse H$\alpha$ emission is low, it is sometime possible to detect  H$\alpha$ counterpart fainter than 10$^{-15}$~erg~s$^{-1}$~cm$^{-2}$. 

We  identify sources in the MIR catalogue with sources in the radio catalogue. For a source at 24~$\mu$m to be identified as the 
MIR counterpart  of a source in the radio catalogue 
we require that the distance between the two sources to be less than the smaller  source radius between that given at radio frequencies  and 
that measured at 24~$\mu$m.  We also visually
analyzed all sources which partially overlap when their distance is less that the largest radius between the radio and MIR one, but not less 
than the smallest of the two, and include some of them in the list of MIR-radio associations. Not all the MIR sources that are left without a radio counterpart represent star 
forming sites with no radio emission because of the following:

$i$)MIR background contamination on source selection

$ii$)MIR source blending and problem with source centroid

$iii$)shift between location of radio emission and that of  hot dust

$iv$)sources lying outside radio maps

$v$)failure of radio source extraction because of faint radio emission contaminated by negative sidelobes.  

We have a radio source counterpart for 330 of the 630 MIR sources and these are plotted in magenta color in the bottom panel of Figure~\ref{RHa_all}.
 In the upper panel we use different colors to mark b-type, c-type and e-type MIR sources associated with radio
sources in the catalogue. We find 7 out of 97 b-type sources, 287 out of 410 c-type sources, and 30 out of 104 e-type sources with  counterpart in the radio catalogue.
As shown in the upper panel of Figure~\ref{RHa_all} the star 
forming regions which have some optical or FUV counterpart (of c-type), show a clear correlation between H$\alpha$ and radio continuum flux.
The red filled circles are for MIR sources associated with known GMCs in \citet{2017A&A...601A.146C} while open symbols are for MIR sources without an
associated GMC because the native cloud is too faint to be detected, either because of its small mass or because  the GMC is dissolving  as the source evolves.  

MIR sources without an optical counterpart do not follow a clear correlation with the radio counterpart but we have to underline that b- and e-type sources
of low 24~$\mu$m flux (below 5~mJy)  suffer of strong background contamination.  Therefore some of the MIR-sources
of b- and especially of e-type with associated radio emission can be background objects \citep{2019A&A...622A.171C}.  
The  e-type  MIR sources have radio spectral indexes much more negative than YSCs,    and distributed around  the value of -0.8; this confirms that they are
a different population, likely associated to background galaxies.  Some MIR sources, which are not in the Figure,
because don't have an associated radio source, might have suffered a displacement of the MIR emission with respect to radio peaks. This can
happen for example as the HII region expands:  the dust can be located just on a side of the ionized bubble while radio emission might peak where
non-thermal or thermal emission is strongest.   On  the other hand radio emission due to embedded compact HII regions  might also have been diluted in the beam. 

\begin{figure}
\hspace{-0.9 cm}
\includegraphics[width=16 cm]{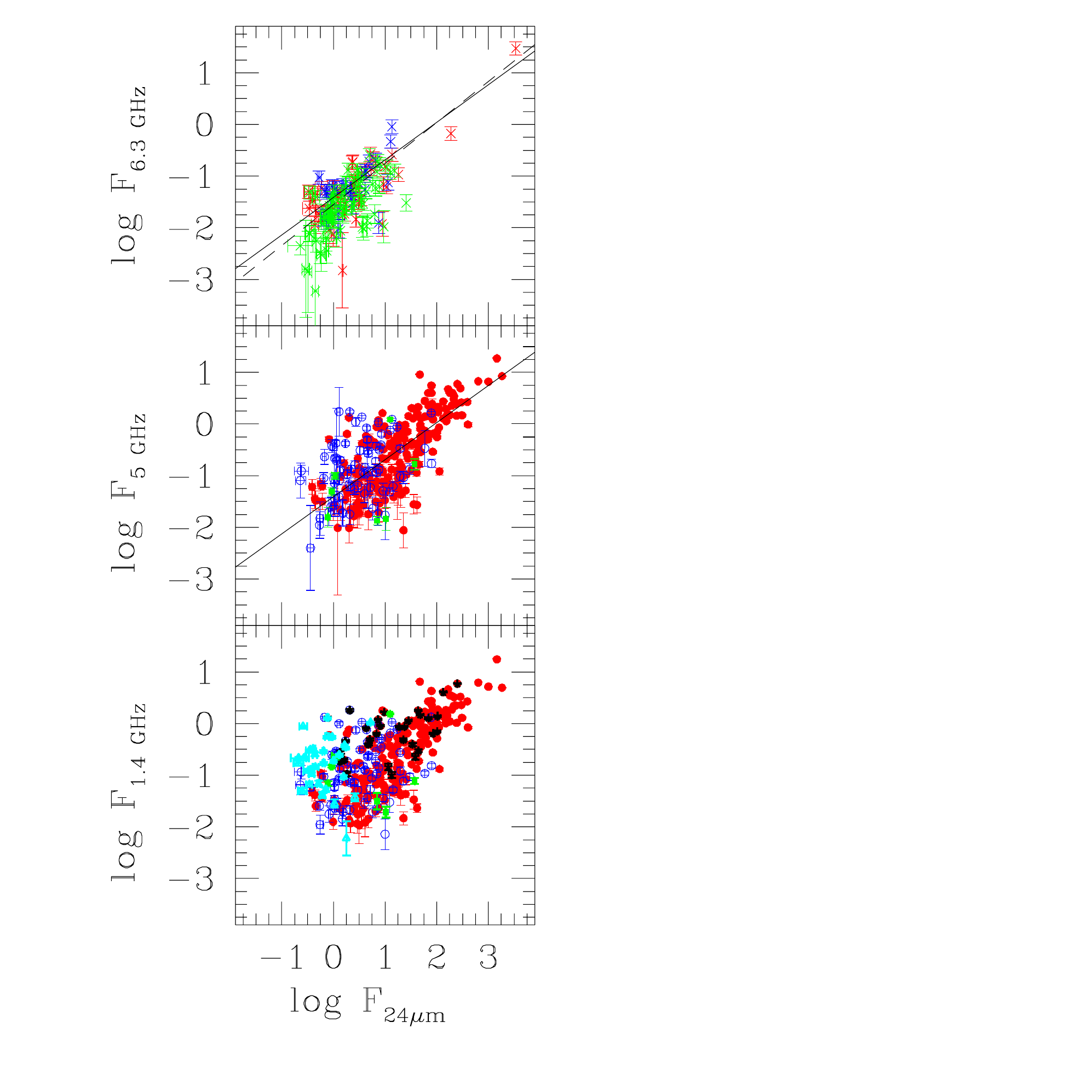}
 \caption{ The bottom panel shows the radio flux at 1.4~GHz of sources in the White et al. catalogue versus the flux at 24~$\mu$m of the associated MIR sources, color coded
 as in Figure~\ref{RHa_all}.  Flux units are mJy. The star symbols indicate the presence of identified SNR in the region (code$>$8).
 In the middle panel  the radio flux at 5~GHz has been computed  using the spectral index of each radio source in the catalogue. 
 We show only data for b-and c-type sources with the continuum  line that is the best fit to them. In the upper panel we display the radio flux at 6.3~GHz from 
 aperture photometry of b- and c-type MIR sources with no catalogued radio counterpart  but covered by the 6.3~GHz survey. 
 Color coding is the same as in the other panels although cross symbols are used for these sources. 
 The continuum line is the same as in the middle panel while dashed line is the fit to all data in the middle and upper panel.
  }
\label{mir}
\end{figure}

In the bottom and middle panel of Figure ~\ref{mir} we plot the MIR flux at 24~$\mu$m versus the radio continuum at 1.4 and 5~GHz  respectively
for all matched catalogued sources, color coded according to their class. 
The 5~GHz flux has been recovered using the 1.4~GHz flux and spectral indexes of \citet{2019ApJS..241...37W}. 
There is a well defined correlation between radio and MIR emission in star forming regions. The correlation gets tighter if we consider sources at $R<3$~kpc
or young sources i.e. located in GMCs (red and green symbols). 
In the  middle panel   the continuous line is the best fit to the distribution which minimizes the geometrical distances to the line i.e. he sum of the squares of 
the distances from the data to the line. The slope of the correlation  shown   is  0.72 and is  unchanged if we replace the 5~GHz with the  6.3~GHz flux.
 The slope is slightly higher, 0.81, if only exposed sources hosted by molecular clouds (red dots) are considered. We would like to underline here that of the 285 HII regions
identified by \citet{2019ApJS..241...37W} only  58 are coincident with MIR sources considered here. We have  in our MIR sample only 31 SNRs (identified
with code 11 or 9)  of the total 104 SNRs with radio continuum peaks.

Radio continuum and mid-infrared radiation are  both linked  to the formation of massive stars and a nearly linear correlation  
is expected when integrated over the whole galaxy extent. The UV radiation from massive stars is partially absorbed by dust  and ionizes the surrounding ISM
where  thermal  electrons  give  rise  to  free-free emission,  while shocks from  winds and SNR  produce cosmic rays that emit  non-thermal (synchrotron) 
radiation. However variations in dust abundance and magnetic field strength justify possible deviations from a linear relation.
Integrated over the whole galaxy     
the average ratio between the 24~$\mu$m and the radio flux density at 1.4~GHz is  lower than what we find in
most star forming regions of M33 \citep{2010ApJ...723.1110H,2004ApJS..154..147A}.  Only  regions hosting a SNR show ratios similar to galaxy integrated values.
This can be attributed to the non continuous production of cosmic rays which diffuse away from  star formation sites,
 but also to the nature of 24$\mu$m emission which is more enhanced where massive stars  form  
 and less diffuse than other infrared tracers.   However, in  M33 dust opacity in star forming regions is not high enough to absorb most of UV and optical starlight  
\citep{2009A&A...493..453V} and 24$\mu$m radiation alone does not trace star formation accurately. The dust abundance, like gas phase metal abundances 
\citep{2010A&A...512A..63M},  decreases with galactocentric distance  and  this variation is also driving the sublinear relation between the 24~$\mu$m emission and
radio continuum.   The lack of strong spiral arms in the  regions beyond corotation  
\citep{2019A&A...622A.171C} favors the birth of small  star forming sites with a lower MIR-to-radio emission ratio than sites in the  inner regions, closer 
to what  is found for galaxies on a larger scale. 
The open blue circles at the faint end of the distribution in Figure~\ref{mir},  are examples of this type of YSCs.  They lie mostly  in the outer disk where CO lines are 
weaker and have no identified  giant molecular clouds.

 \subsection{Radio continuum photometry of undetected counterparts} 

We now consider all the MIR sources which are associated with a GMC or that have an optical counterpart and very likely are  star forming regions
in the M33 disk (we refer to these as Young Stellar Clusters, hereafter YSCs). Those are listed as sources type-b,-c,-d in the
classification scheme of \citet{2017A&A...601A.146C} (see their Table 2) and are in total 526, 300 of which have an associated radio source.
To complement the existing radio-catalogue by deeper searches for diffuse radio-emission associated with the remaining MIR sources, we carried out
aperture photometry on the radio map at 6.3~GHz. This covers a smaller region than the 5 and 1.4~GHz maps and hence a substantial
fraction (about 30$\%$ of the total )  of the MIR sample lies outside this map.   

Of the sources inside the map which had no radio counterpart in the catalogue we detect 90$\%$ of them using aperture photometry. 
 By adding the number of  detected MIR sources with aperture photometry to
the number of matched sources listed in catalogues we have a radio counterpart for 440 MIR sources.   
Inside the 6.3~GHz map we can claim a detection rate of 95$\%$  including catalogued and non-catalogued radio counterparts.  
Given the fact that the non-detected sources are small and affected by beam dilution, and moreover some might have had a bright radio
source nearby that  artificially affected the estimated background, we can say that MIR-source located in the inner disk 
(at R$< 4$~kpc)   and related to star-forming regions have associated radio-emission. 
The fraction of MIR sources with a radio counterpart at larger radii is $> 60\%$ but we cannot constrain this number better due to the limited coverage
of the 6.3~GHz map. 
 
In the upper panel of Figure~\ref{mir} we show the radio flux recovered by aperture photometry on the 6.3~GHz map,
versus the 24~$\mu$m  flux of MIR star forming sources with no catalogued radio counterpart. The
dashed line is the fit to all sources in the middle and upper panel of the Figure. Its slope is 0.80 slightly higher than the fit relative to matched catalogued sources only.
 Clearly YSCs without a catalogued radio counterpart
have weaker radio fluxes than  YSCs with similar MIR fluxes. This is also because most of these sources are still embedded, in the early phases of star formation prior to 
 the onset of winds and shocks  with enhanced turbulent magnetic fields and cosmic ray acceleration. However, there is a good agreement of the radio fluxes recovered
via aperture photometry at 6.3~GHz with those recovered by source extraction at slightly lower frequency. This proves
that indeed there is only a unique population of YSCs in M33 and that the relation between the radio-continuum at 6.3~GHz and the
MIR flux is slightly sublinear, with the MIR flux increasing faster than the radio flux does as the YSCs becomes brighter.

\section{Thermal and non-thermal radio emission} 

In this Section we investigate  the fraction of radio emission that is linked to thermal and non-thermal processes in YSCs. To derive the
thermal fraction in YSCs we use the extinction corrected H$\alpha$ flux.
This has been computed  using  the expression of the free-free absorption coefficient   in the Rayleigh-Jeans regime with the velocity averaged
Gaunt factor given by \citet{2011piim.book.....D} and assuming that hydrogen atoms provides most of the ions and free electrons:

\begin{equation}
{k_\nu\over \hbox{cm}^{-1}} = 1.091 n_e n_i 10^{-25} ({T\over 10^4 K})^{-1.32} ({\nu \over 1 GHz})^{-2.118} 
\end{equation}

where $\nu$ is the frequency in GHz, T is the electron temperature in K, n$_e$
and n$_i$ are the electron and ion densities in cm$^{-3}$. The brightness temperature is the product between the electron temperature and the optical depth $\tau_\nu$,  
the latter being  the integration of $k_\nu$ along the line of sight. Using the emission measure EM in cm$^{-6}$~pc we can write the brightness temperature    for
free-free emission as:

  \begin{equation}
 {T_b^{ff}\over \hbox{K}} \simeq \tau_\nu T = 3.37\ 10^{-3}\ ({T\over 10^4 K})^{-0.323}\ ({\nu \over 1GHz})^{-2.118} ({EM \over cm^{-6} pc})   \\
   \end{equation}
 
\noindent
The flux density for a source of uniform brightness is

\begin{equation}
S_\nu^{ff} = {2\nu^2 k_B T_b^{ff} \over c^2} \Omega_s
\end{equation}

with k$_B$ the Boltzmann's constant, c the speed of light and $\Omega_s$ the angular source extent. Using the above expression for the brightness temperature the
flux density in milliJansky is

\begin{equation}
{S_\nu^{ff} \over  mJy} = 0.1033  ({T\over 10^4 K})^{-0.323}\ ({\nu \over 1GHz})^{-0.118} ({EM \over cm^{-6} pc})  
\end{equation}

\noindent
The  intensity of H$\alpha$ line emission  \citet{2011ApJ...727...35D} is
 
 \begin{equation}
{ I_{H\alpha}\over  \hbox{ergs}~\hbox{cm}^{-2}~\hbox{s}^{-1}~\hbox{sr}^{-1}} = 0.87\ 10^{-7}\  ({T\over 10^4 K})^{-0.94-0.031 ln(T/10^4)}\  ({EM \over cm^{-6} pc}) 
  \end{equation}
 
 \noindent
 For recovering F$_{H\alpha}$, the H$\alpha$ flux over the source extent, we multiply the above expression by $\Omega_s$.
 We can then write the  free-free flux density in milliJansky as a function of   F$_{H\alpha}$ in ergs~cm$^{-2}$~s$^{-1}$ as
 
 \begin{equation}
 {S_\nu^{ff} \over  mJy}= 1.187 \  {F_{H\alpha}\over  10^{-12} \hbox{ergs}~\hbox{cm}^{-2}~\hbox{s}^{-1}}\ ({\nu \over 1GHz})^{-0.118} ({T\over 10^4 K})^{y(T)}
  \end{equation}
  
 \noindent
 where we have defined y(T)={0.619+0.031 ln(T/10$^4$)}.
   
\begin{figure*}
\hspace{-1.2 cm}
\vspace{-10.2 cm}
 \includegraphics[width=18.5 cm]{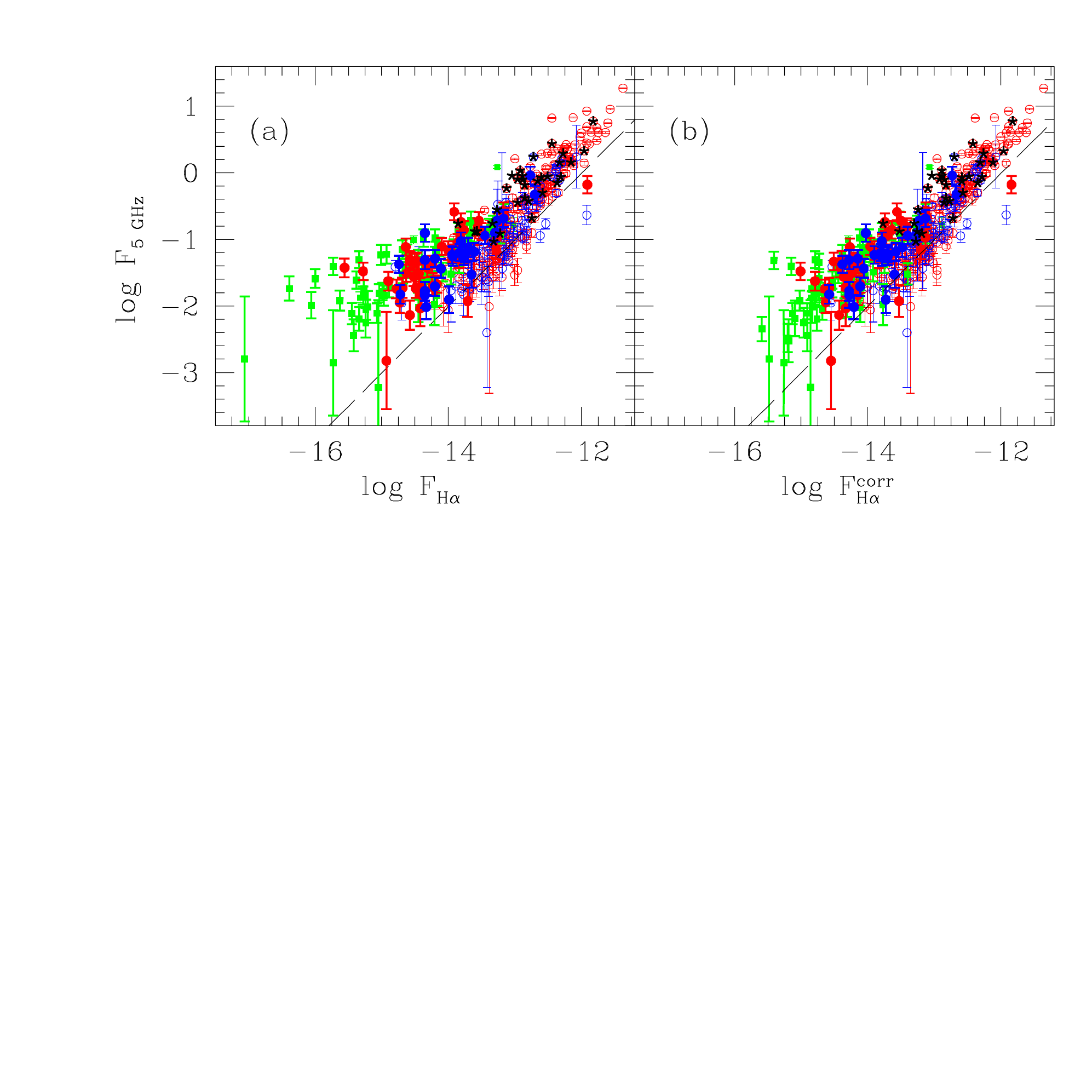}
 \caption{The H$\alpha$ fluxes in erg cm$^{-2}$ s$^{-1}$ measured  through aperture photometry at the location of  radio sources associated with MIR emission are
 shown in panel $(a)$  as a function  of the radio flux  recovered from catalogued data at  at 5~GHz  in mJy. 
 Color coding is the same as in Figure~\ref{RHa_all} but open symbols have been used.  
 In the right panel $(b)$  we display for the same sources the H$\alpha$ flux corrected for internal extinction. Filled symbols indicates radio
 fluxes from aperture photometry at 6.3~GHz of b- and c-type MIR sources with no catalogued radio counterpart. 
 The predicted H$\alpha$ emission if the radio flux were due to free-free radiation only is shown with a dashed line. The star symbols indicate the presence of identified 
 SNR in the regions (catalogue code$>$8). 
  }
\label{hacorr}
\end{figure*}

From now on we shall use the H$\alpha$ flux values measured by our aperture photometry centered on radio sources which are in good 
 agreement with those of   \citet{2011A&A...534A..96S} centered on MIR sources. 
 The correlation between H$\alpha$ fluxes, from our aperture photometry measurements, and  fluxes at 5~Ghz for radio sources associated to YSCs are
shown in  Figure~\ref{hacorr}. The dashed line indicates the expected H$\alpha$ flux if the radio emission were due to free-free emission only.
 This has been recovered using eq.(6) for an average electron temperature of 10$^4$~K across the star forming disk of M33. 
 In the left panels we have not  corrected the H$\alpha$ line for internal extinction (the Galactic extinction is negligible in the direction of M33).  
  
We use the MIR emission to correct the H$\alpha$ for internal extinction according to   \citet{2009ApJ...703.1672K}  as follows:

\begin{equation}
F_{H\alpha}^{corr}= F_{H\alpha}  + 0.02 \nu_{24} {F_{24}\over erg~cm^{-2}~s^{-1}} C 
\end{equation}

\begin{equation}
{F_{H\alpha}^{corr}\over 10^{-15}} = {F_{H\alpha}\over 10^{-15}}  + 2.5 {F_{24}\over mJy} C
\end{equation}

where the H$\alpha$ flux is in erg~s$^{-1}$~cm$^{-2}$ and the observed value is F$_{H\alpha}$.
 C is a correction factor which takes into account the different   MIR and radio source radius. 
 The extent of MIR sources is usually larger than the associated  radio sources for large fluxes and therefore we correct the  24$\mu$m  flux values given by 
\citet{2011A&A...534A..96S} for the different source extent using the factor C equal to the ratio of the radio-to-MIR  source area.  
 In the right panel of Figure~\ref{hacorr} we show   the extinction corrected   H$\alpha$ flux associated to YSCs. These corrections are especially important for 
 MIR sources associated with GMCs since extinction corrections here are larger, although for M33 these are
 never extremes. In Figure~\ref{hacorr}  we add  the 6.3~GHz  radio and H$\alpha$ fluxes   for MIR sources without radio catalogued 
 counterpart for which the radio flux at 6.3~GHz has been recovered by aperture photometry. A comparison between the left and right panels in Figure~\ref{hacorr}
 shows that extinction corrections are particularly relevant  for embedded YSCs. The e-type sources, not shown in the Figure, do not follow the general correlation and 
 this confirms that these are likely background galaxies.
 The  Figure underlines again that small and  dimmer YSCs in M33 have both radio continuum and H$\alpha$
 emission, although  resolution and sensitivity limits imply that often these cannot be recover by using source extraction algorithms.
 The relation is  in agreement with that found for brighter sources and hence we conclude that they belong to the same population.

The correlation between H$\alpha$ and radio fluxes extend over 4 orders of magnitudes. The expected thermal radio emission computed from the H$\alpha$ 
emission using equation (6)  is shown with a dashed line also in panel $(b)$.
The radio flux associated with star forming regions is  2-2.5 times stronger on average than expected if only   thermal radio emission were associated to the YSCs.
We notice the unambiguous presence of non-thermal emission in MIR sources with H$\alpha$ flux greater than $2\times10^{-13}$~erg~s$^{-1}$~cm$^{-2}$ which correspond to a
luminosity of  2$\times10^{37}$~erg~s$^{-1}$ at the distance of M33, equivalent to  a cluster populated up to O7-type star. Indeed theoretical computations predicts that only O-type stars
more massive than M* are able to produce a wind and therefore a shock. Despite the value M* is still debated  it is close to that of O7-O6 spectral type
\citep{2000A&A...362..295V,2009A&A...498..837M, 2012A&A...537A..37M}.

\subsection{Possible dependencies of non-thermal fractions on source compactness and  galactocentric distance}
 
 Below we refine our computation to investigate the variations of thermal and non-thermal fractions with the luminosity, age, and local environment of the star forming
regions. The electron temperature, needed to estimate the free-free emission,  can be measured in HII regions through  spectroscopy
of  auroral lines.   For  M33   \citet{2010A&A...512A..63M}
have derived  how  this temperature increases with galactocentric radius. Expressing the latter in kpc  the radial dependence of the electron temperature in M33 reads:

 \begin{equation}
 \frac{T_e}{\rm K} = 410 \frac{R}{\rm kpc} + 8600
 \end{equation}
 
 From now on we measure the total radio flux and the corresponding H$\alpha$ flux by  performing aperture photometry both on the 5~GHz map and on the H$\alpha$ map, the latter
  smoothed to the resolution of the radio map. We consider  only
  YSCs of c-type i.e. MIR sources with radio counterpart in the catalogue of \citet{2019ApJS..241...37W} which are no longer fully embedded into the native molecular cloud 
  but  have an optical or UV counterpart.
   We divide these YSCs into 3 categories according to their radius R$_s$: $small$ for 3$\le$R$_s < 6$~ arcsec,  $medium$ for $6\le R_s < 9$~ arcsec, and $large$ for 
   9$\le$R$_s < 18$~ arcsec. The radius R$_s$ is the average between source semiminor and semimajor axes at 1.4~GHz as listed in the  catalogue of \citet{2019ApJS..241...37W}. 
    The non-thermal component is just the difference between the total radio flux and the thermal component, the latter being computed using equation (6) with the radially 
    dependent electron temperature T$_e$.  To check that the resulting non-thermal fractions do not depend much on the aperture,  we vary the aperture  size 
    but use consistently the same aperture on both maps and compute both backgrounds in  annuli placed at distances of 3-3.5 times the source radius.
    
\begin{figure}
\includegraphics[width=9.0 cm]{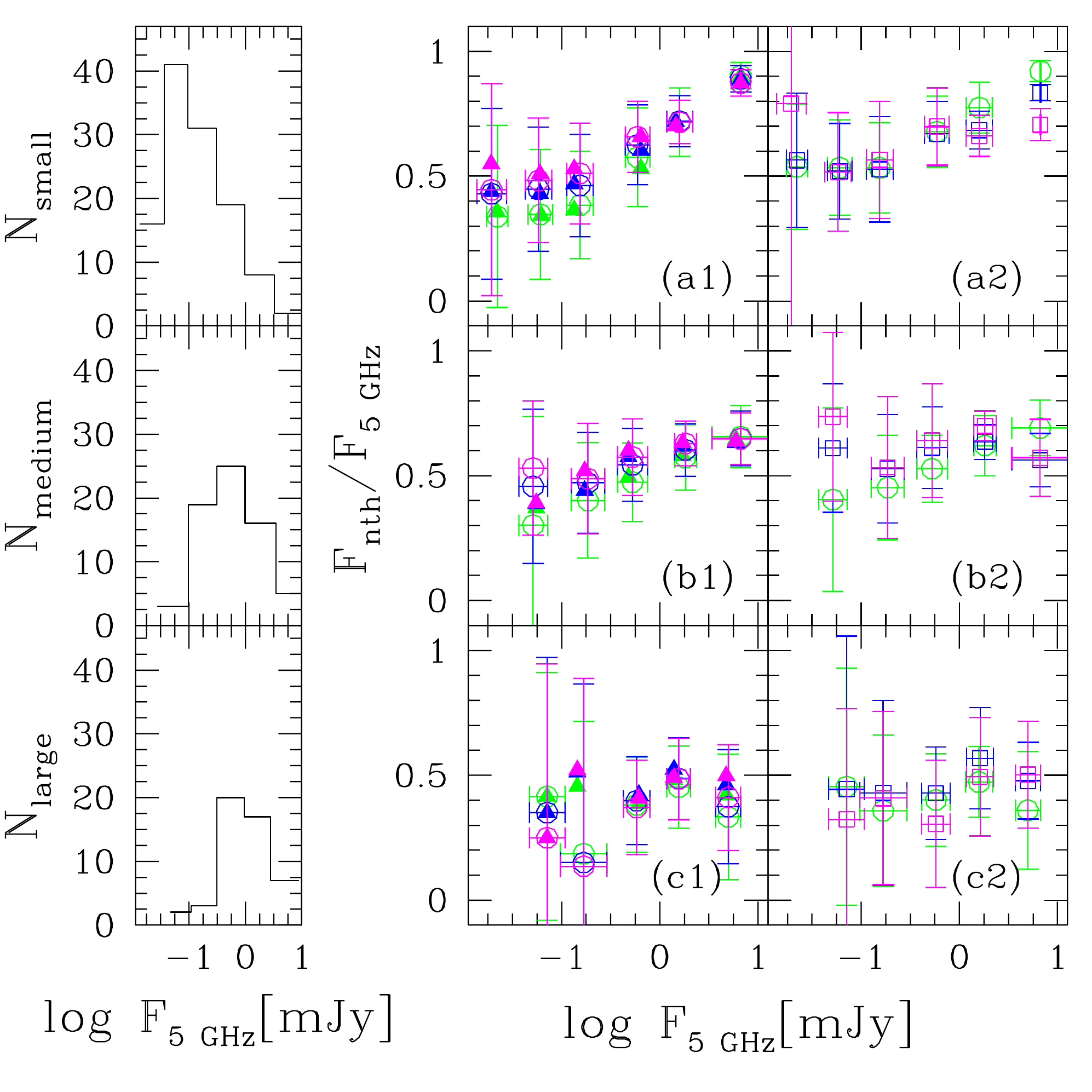}
 \caption{The non-thermal fraction  at  5~GHz of radio sources associated with MIR sources in exposed HII regions is shown in  $(a1)$ for small sources,
 in $(b1)$ for medium sources and in $(c1)$ for large sources as a function of the  radio flux density at the same frequency.   
 Filled triangles are the median values, open circles the mean values with the standard deviations. Colors indicates different apertures: 
 green for R$_{ap}$=R$_s$, blue for R$_{ap}$=1.5 R$_s$ and magenta for R$_{ap}$=2 R$_s$.  Left panels show the number of sources in each bin for sources of  small  
 (top),  medium (middle) and large (bottom) size. Right panels 
 display the non-thermal fractions  for  R$_{ap}$=R$_s$ (open green circles), in the annulus between R$_s$ and 1.5R$_s$ (open blue squares) and between 1.5R$_s$
 and 2R$_s$ (open magenta squares) for small, medium and large sources ($a2,b2,c2$ respectively).  Radio flux units are mJy. }
\label{size}
\end{figure}

 Figure~\ref{size} shows the resulting non-thermal to total radio flux ratio as a function of the 5~GHz radio flux density F$_{5GHz}$. This is the total flux density in mJy 
in a circular aperture whose radius is R$_{ap}$=1.5R$_s$.  Small, medium and large sources are displayed in the top, middle and bottom rows respectively.
The histograms to the left  show the number of  sources in each flux bin. The central panel  shows the median (filled triangles) and the mean (open circles) non-thermal fractions
with their  standard deviations  for  each flux bin. The green, blue and magenta colors indicate  apertures with radii of 1,1.5 and 2~R$_s$ respectively. 
We  see  that there is a common trend for all apertures for both the mean and median values. We have also checked our results using photometric fluxes with no
background subtraction.  Large sources show no clear dependence of the non-thermal fraction with radio flux density and have lower  fractions than more compact
sources of the same luminosity.  An increase of the mean and median non-thermal fractions as a function of radio flux density are clearly seen for medium  and small source sizes. 
The most compact sources brighter than 0.3~mJy show the steepest rise with radio flux density.  These
sources  do not host catalogued SNRs  but  have non-thermal fractions $>50\%$ at 5~GHz.  Faint radio sources in low mass YSCs have a large 
dispersion around the mean  because of the stochastic character of the Initial Mass Function \citep{ 2009A&A...495..479C} which implies that  low mass YSCs 
only occasionally host massive stars. 
Star forming regions with a large non-thermal to thermal radio flux ratio follow the high surface density knots of gaseous filaments while the others are more coarsely placed
around or on the filaments.

In the last panel of Figure~\ref{size} we plot again the mean non-thermal
fraction in apertures with R$_{ap}$= R$_s$ (green color). We add to these points the mean non-thermal fraction in the annulus between 1 
and 1.5~R$_s$ (blue color) and   between 1.5 and 2~R$_s$ (magenta color). We can see, by comparing panel $(a1)$ and $(a2)$, that the excess 
of non-thermal flux for compact bright sources
is close to the peak of the emission (green circles) since the external annuli (in magenta) have lower non-thermal fractions. There is no  consistent  monotonic trend  for
 variations of  non-thermal fractions with distance from the radio emission peaks, and no obvious correlation with the radio spectral index given in the catalogue.

\begin{figure}
\hspace{-0.5 cm}
\includegraphics[width=12.5 cm]{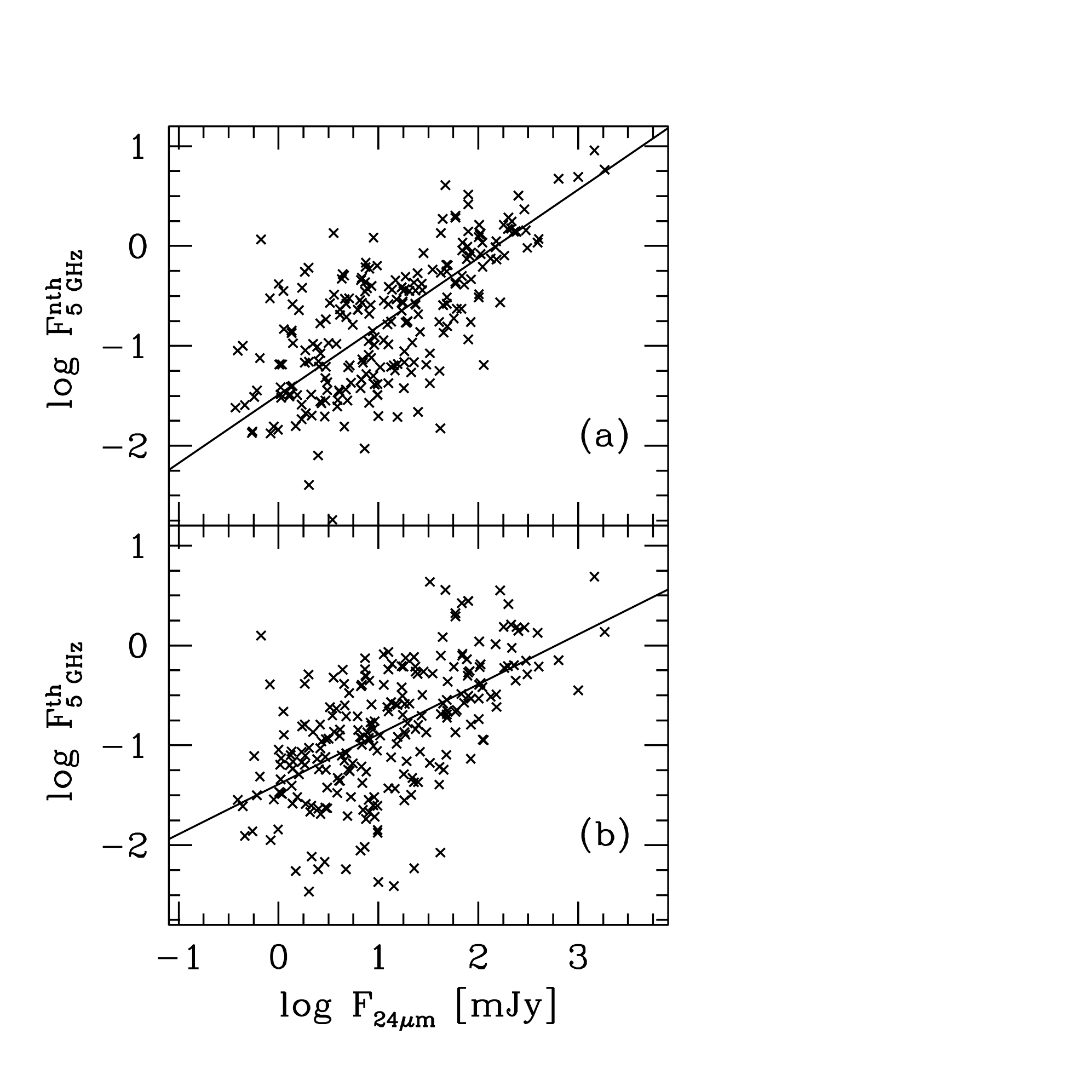}
 \caption{Correlations between 
 the   non thermal $(a)$ and thermal $(b)$ radio continuum at 5~GHz with the 24~$\mu$m emission in 248 exposed HII regions of M33. 
 The lines minimize distances of data points to them and have slopes of   
 0.68 and 0.50 in panel (a) and (b) respectively. Radio flux units are mJy and have been measured using circular apertures with R$_{ap}$=1.5 R$_s$. }
\label{nth}
\end{figure}

In the  two panels of Figure~\ref{nth} we show the correlations between the thermal and non thermal radio continuum at 5~GHz with the 24~$\mu$m emission.
The linear relations in the log-log plane have slopes of 0.50 and 0.68 for the  thermal and non thermal radio emission respectively (by minimizing distances 
of  data points to a straight line). The larger scatter at low 
luminosities  in the non-thermal radio continuum-MIR correlation is  likely due to non-continuous injections of fast particles and to   stochastic sampling of  
the upper end of the stellar IMF (Initial Mass Function)  which settles in as the cluster mass decreases \citep{2009A&A...495..479C}. 
Turbulent magnetic field amplification and CR acceleration clearly depend on the presence of shocks related to massive stars and might be enhanced
in some low luminosity star forming regions that host massive star outliers. The slope of the relation is  similar to what \citet{2017MNRAS.471..337B}  
find sampling IC10 at 6.2~GHz  on 55~pc scale. The thermal radio continuum declines more slowly with the YSC MIR luminosity 
than the non-thermal component does.   
The  MIR sources are not overlapping completely with  their radio continuum counterparts and in particular they are larger for bright sources. 
This implies that the total H$\alpha$ flux corresponding to the MIR source, and hence the thermal radio continuum,  might be
larger in the whole star forming region what has been used  in  Figure~\ref{nth} and which refers to the location of the radio source.  
An additional  diffuse radio thermal component might be present for bright sources which would make the MIR-thermal radio continuum correlation steeper. 
This can be checked in the future at larger scales using  radio maps corrected for short spacing.

\begin{figure}
\hspace{-1.5 cm}
\includegraphics[width=12. cm]{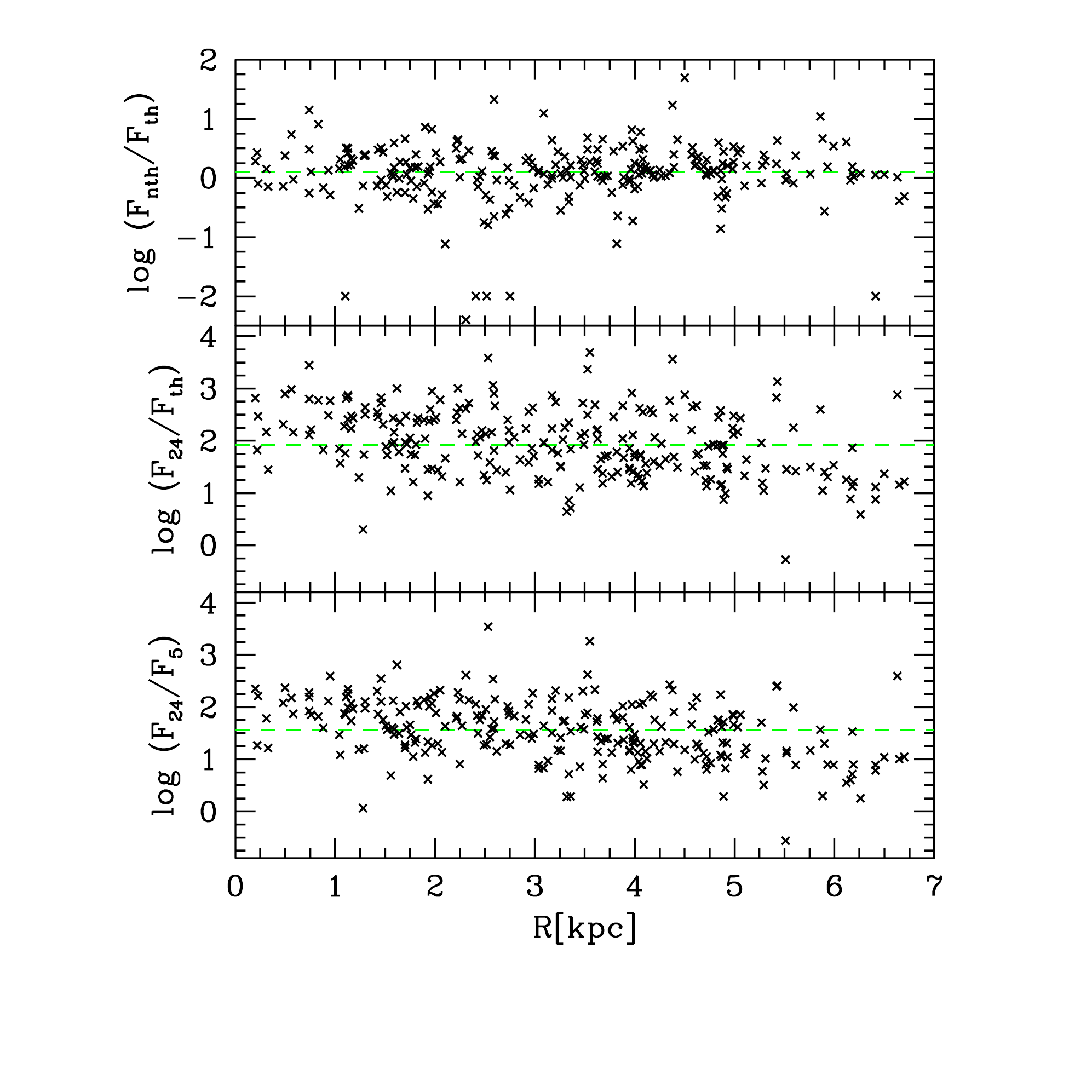}
 \caption{The radial trend of MIR-to-radio flux ratio at 5~GHz (bottom), for the thermal radio emission only (middle), and 
 the non-thermal-to-thermal ratio  at  5~GHz (top) are plotted for exposed YSCs with radio counterparts in the catalogue.
  Dashed green lines are reference lines placed at the median value of the distributions.
  }
\label{radial}
\end{figure}

In Figure~\ref{radial} we show the galactocentric radial dependences of the MIR-to-radio continuum and MIR-to-thermal radio continuum and non-thermal-to-thermal radio 
continuum ratios. The green dashed lines are plotted for reference and indicate the median value of the distributions.
The radial slopes  in the bottom 2 panels of Figure~\ref{radial} are similar and   driven 
by the radial decrease of dust abundance, particularly evident beyond  3.5~kpc. Sources with F$_{th}$/F$_{nth}>1$ at 5~GHz  are 40$\%$ of the total number of sources
at R$<3.5$~kpc. This percentage decreases to 25$\%$ for R$>3.5$~kpc. Non-thermal emission therefore increases with galactocentric distance although the
electron temperature in HII regions increases as well to power thermal emission. This result is consistent  with a  decrease in the radio thermal fraction    on a larger scale
going  radially outwards in M33 \citep{2007A&A...472..785T}.

\section{Radio emission from molecular clouds}

\begin{figure} 
\centerline{
\includegraphics[width=8cm]{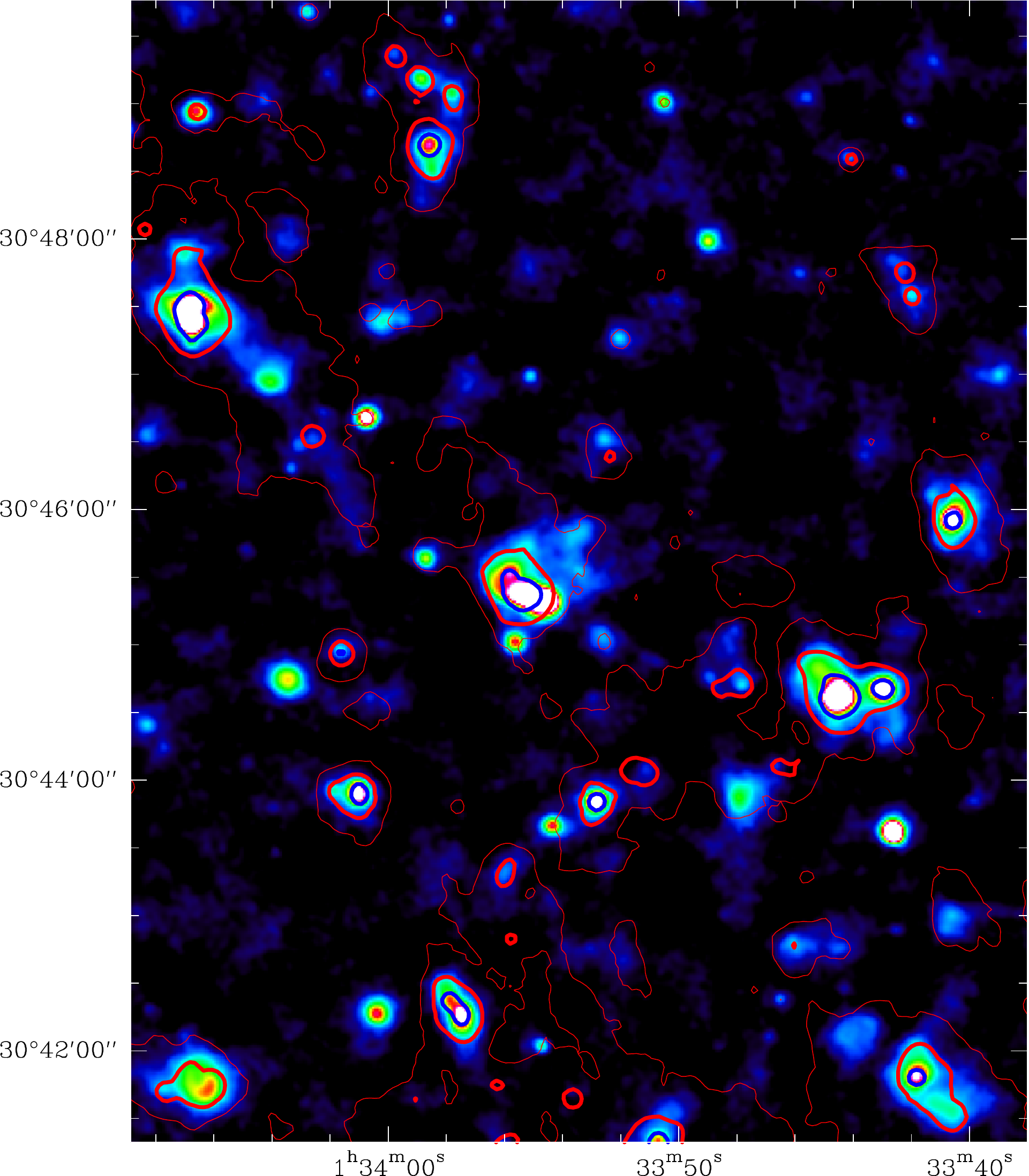}}
\includegraphics[width=9cm]{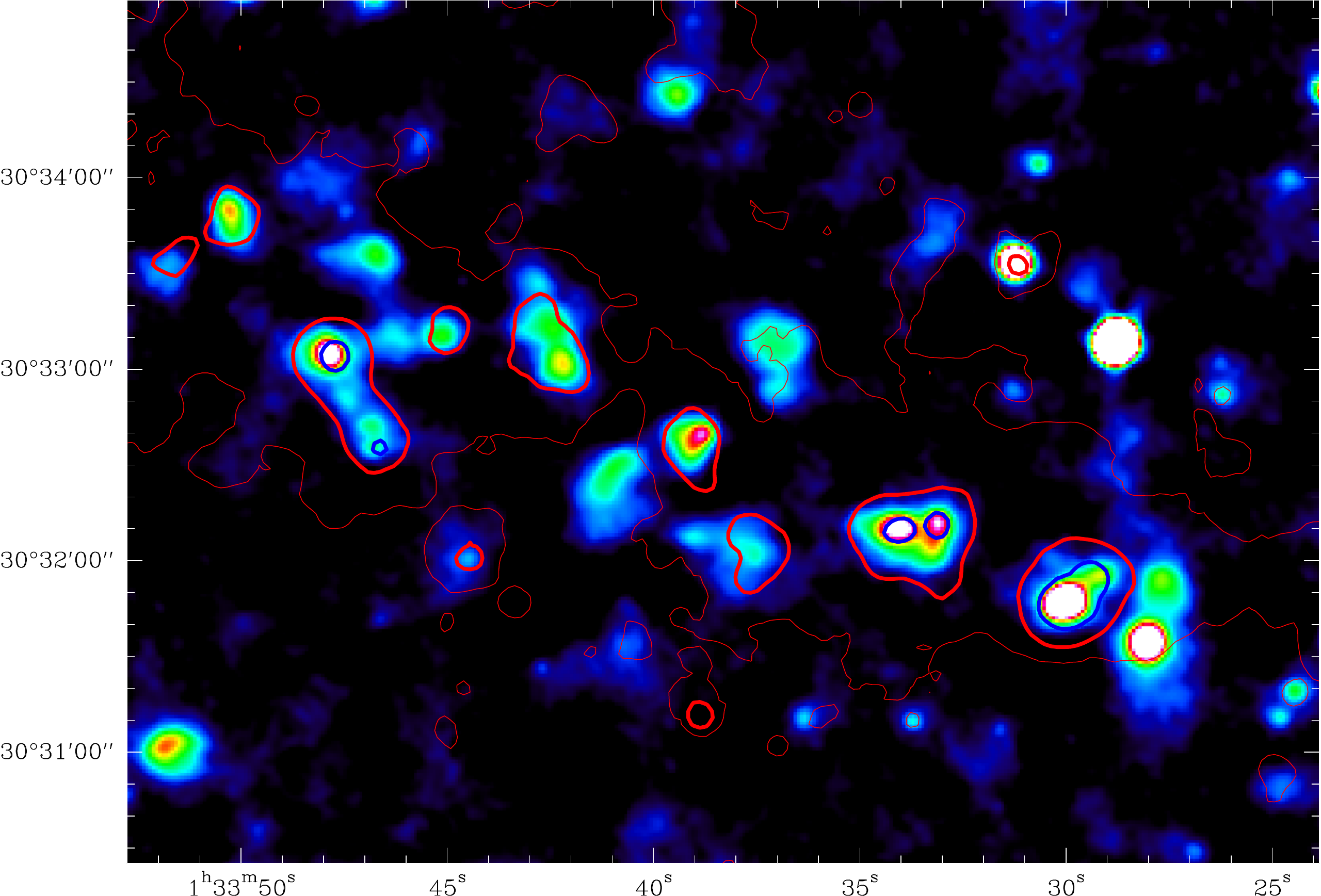}
\caption{We compare radio and MIR emission of a kpc-size region allowing details of tens of parsecs in size to be seen.
Color image is 1.4 GHz radio emission from  \citet{2019ApJS..241...37W}  shown with a log scale of a portion of the northern (upper panel ) and southern (bottom panel) 
side of M33.  Contours are 24 micron Spitzer emission  at 1, 3, and 10 (thick blue line) MJy/sr.   
}
\label{radiomir} 
\end{figure}

One of the most striking aspects of the radio emission is how it follows the star formation.
This is true for both the thermal and non-thermal emission. In Figure~\ref{radiomir} we compare the radio emission at 1.4~GHz with the
thermal dust emission at 24$\mu$m in two kpc-size arm regions, one in the northern side and the other on the southern side.

Zooming into 100~pc size regions we can investigate the radio emission from individual molecular clouds and how this relates to other
tracers of star formation.
Each molecular cloud covers a well-defined region in the sky (e.g. Fig.4 of \citet{2017A&A...601A.146C}) and we now discuss the radio emission from the GMC sample, using 
photometry of the regions within the cloud boundaries. We  exclude for the statistical analysis  clouds outside the 5~GHz map and in the proximity of NGC604. 

\begin{table*}
\caption{Median and mean radio fluxes at 5~GHz, H$\alpha$ and 24$\mu$m emission for each GMC class} 
\centering                                       
\begin{tabular}{c c c c c c c c c c c}           
\hline\hline 
GMC  &   GMC  &F$^{GMC}_{5}$&F$^{GMC}_{5}$-F$^{srcs}_5$& F$^{GMC}_{H\alpha}$    & F$^{GMC}_{24}$  & < M$^{GMC}$> & <F$^{GMC}_{5}$> & <F$^{GMC}_{5}$-F$^{srcs}_5$> & <F$^{GMC}_{H\alpha}$> & <F$^{GMC}_{24}$>\\   
type   &       N       &$\mu$Jy             & $\mu$Jy                                        &ergs~cm$^{-2}$~s$^{-1}$  & $\mu$Jy                  & M$_\odot$         &$\mu$Jy                   & $\mu$Jy                                             &ergs~cm$^{-2}$~s$^{-1}$  & mJy \\ 
\hline\hline 
A   &  169   &  -7.   &    -7.   & 7.2 10$^{-15}$  & 4. &   1.3 10$^5$  &  -2.   &  -6. & 1.4 10$^{-14}$   & 6. \\
B   &   86   &  20.    &   11.   & 1.5 10$^{-14}$  & 9. &    2.1 10$^5$ &  52.   &   7.    &    2.8 10$^{-14}$  & 13. \\
C   &   276  & 282. &   85.   &  6.7 10$^{-14}$ & 19. &    3.6 10$^5$  &  795. &   251. &    1.5 10$^{-13}$ & 48. \\
 \hline\hline 
\end{tabular}
\label{tbl:fitresults}
\end{table*}

We compute both the total radio emission and the residual emission for each GMC. Residual emission has been estimated after subtracting  
the flux of radio sources in the catalogue of \citet{2019ApJS..241...37W} 
associated with each GMC,  F$^{GMC}$-F$^{srcs}$, considering radio sources within the average cloud radius. For individual clouds
the residual emission can be negative if, for example, the radio source is close to cloud boundary because some flux might not be within the GMC contour. 
Table~1 lists for each cloud class the median (F$^{GMC}_5$) and mean (<F$^{GMC}_5$>) fluxes at 5~GHz, the median and mean residual emission (F$^{GMC}_5$-F$^{srcs}_5$
and <F$^{GMC}_5$-F$^{srcs}_5$> respectively). In addition we show the mean mass of GMCs, the median and mean H$\alpha$ and 24$\mu$m fluxes for each cloud class.
Radio emission is clearly detected for most of the clouds with exposed star formation (C class) with  a wide dispersion: fluxes are distributed mostly between the values of
50 and 5000~$\mu$Jy at 5~GHz with a median value of 282~$\mu$Jy.  Radio emission at 5~GHz  is  detected  for about  80~$\%$ of  B-type clouds with most of the fluxes being 
between 50 and 500~$\mu$Jy and a median value of 20~$\mu$Jy, much lower than for C-type clouds. For inactive clouds, 
or GMCs without massive star formation (A class), radio fluxes are distributed around the value of zero. 

The low detection rate of clouds without star formation is 
interesting but requires further investigation using radio maps corrected for missing short spacing. Molecular clouds are expected to concentrate magnetic fields, 
which could result in some diffuse non-thermal emission if cosmic rays penetrate the clouds.  There are 3 A-type 
clouds next to NGC604 which show enhanced  radio emission and  a double peak in the CO J=1-0 line. Likely strong shocks in the expanding shell of NGC604 compress the 
surrounding interstellar medium favoring the formation of GMCs and the propagation of relativistic particles streaming along the magnetic field lines.
In a future paper, dedicated to the mechanism for cosmic ray acceleration, we will use radio maps corrected for short spacing to examine   in more detail  
diffuse radio emission in GMCs at various stages of evolution and at different locations in the disk. 

\begin{figure}
\hspace{-0.5 cm}
\includegraphics[width=9.5 cm]{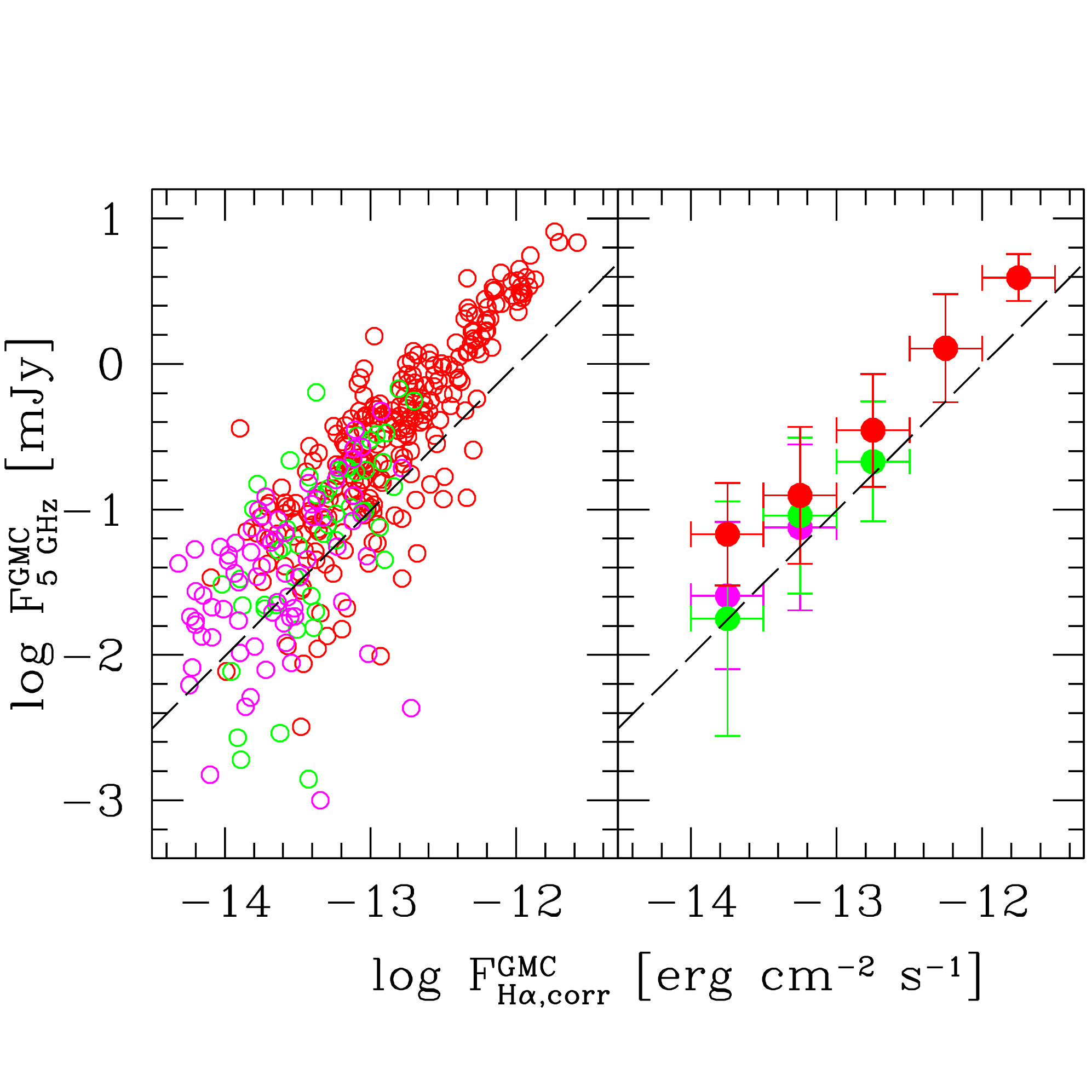}
 \caption{The radio at  5~GHz versus the H$\alpha$ emission for each GMC is displayed in the left panel. In the right panel the mean  radio emission from GMCs in bins of  H$\alpha$  is shown.  
 An average correction for extinction,
using the 24~$\mu$m emission associated with each cloud, has been applied to the H$\alpha$ flux of individual GMCs. Red circles are relative to C-type clouds with exposed 
YSCs, green circles are for B-type clouds with embedded star formation and magenta circles   for inactive clouds of A-type. Only data for GMCs detected at 5~GHz have been used. 
The dashed line shows the expected thermal component at 5~GHz}
\label{clouds}
\end{figure}

Residual emission, not associated with identified radio sources, is barely present in B-type clouds  but it represents about 1/3 of the total radio emission,
for GMCs of class C, with exposed star formation.  As star formation  progresses, and the GMC becomes of  C-type with the YSC   breaking through the cloud, 
the  increase of radio emission in GMCs 
is much stronger than any corresponding increase of thermal emission (as indicated by H$\alpha$ recombination lines or hot dust emission).  As shown in Table~1, in fact, 
both  H$\alpha$  or 24$\mu$m emission increase only by a few from B to C class clouds.
The average radio emission instead increases by more than one order of magnitude suggesting a substantial increase of the non-thermal component
as the YSC evolves. In Figure~\ref{clouds} we show the correlation between the radio and the H$\alpha$ flux from GMCs; in the right panel we have binned the GMCs according to 
their H$\alpha$ emission. An average correction for extinction has been applied to the H$\alpha$ emission from GMCs using Eq.(8), where all fluxes refer to integrated quantities 
inside each GMC contour. The trend for the red dots, which indicate
clouds with exposed star formation (C-type), is very similar to what we found for individual YSCs. The Figure underlines the weaker radio emission of A- and B-type GMCs (magenta
and green dots respectively).
The large dispersion, as indicated by the errorbars, is due both to variations of the star formation rate in GMCs and to some  sidelobe affecting the 5~GHz map.

\section{The link between radio emission and star formation: from GMCs to YSC scale }

We have seen in the previous Section that thermal radio emission in HII regions can be estimated from recombination lines  such as  H$\alpha$  
and we have used the 24~$\mu$m flux density  to correct H$\alpha$  for internal extinction (generally low for exposed YSCs in M33).
Extinction corrected H$\alpha$ emission  provides an estimate not only of the thermal radio flux density but also of the local star formation rate. 
For the scales we are examining here, from GMCs to YSCs, star formation is an event  that lasts less than the GMC lifetime, the latter being estimated of order
14~Myrs \citep{2017A&A...601A.146C}. YSCs start breaking through the cloud when they are about 5~Myrs old, after the embedded phase, \citep{2017A&A...601A.146C}.
Hence, the timescales involved when sampling GMCs and YSCs are of order 10~Myrs, until the GMC disperses through the ISM.
If the stellar mass of the newly formed cluster is not sufficient  to fully populate the IMF at its upper end, the H$\alpha$ is no longer a good star formation rate indicator. 
This affects  cluster masses below  10$^3$~M$_\odot$ or H$\alpha$ luminosities  lower  than  10$^{37}$~erg~s$^{-1}$,
corresponding to a flux of about 10$^{-13}$~erg~cm$^{-2}$~s$^{-1}$ at the distance of M33 \citep{2011A&A...534A..96S}.  At lower luminosities stochastic sampling of the IMF 
with occasional formation of massive stars implies one should use mean values of ensembles of sources.

\begin{figure}
\includegraphics[width=9. cm]{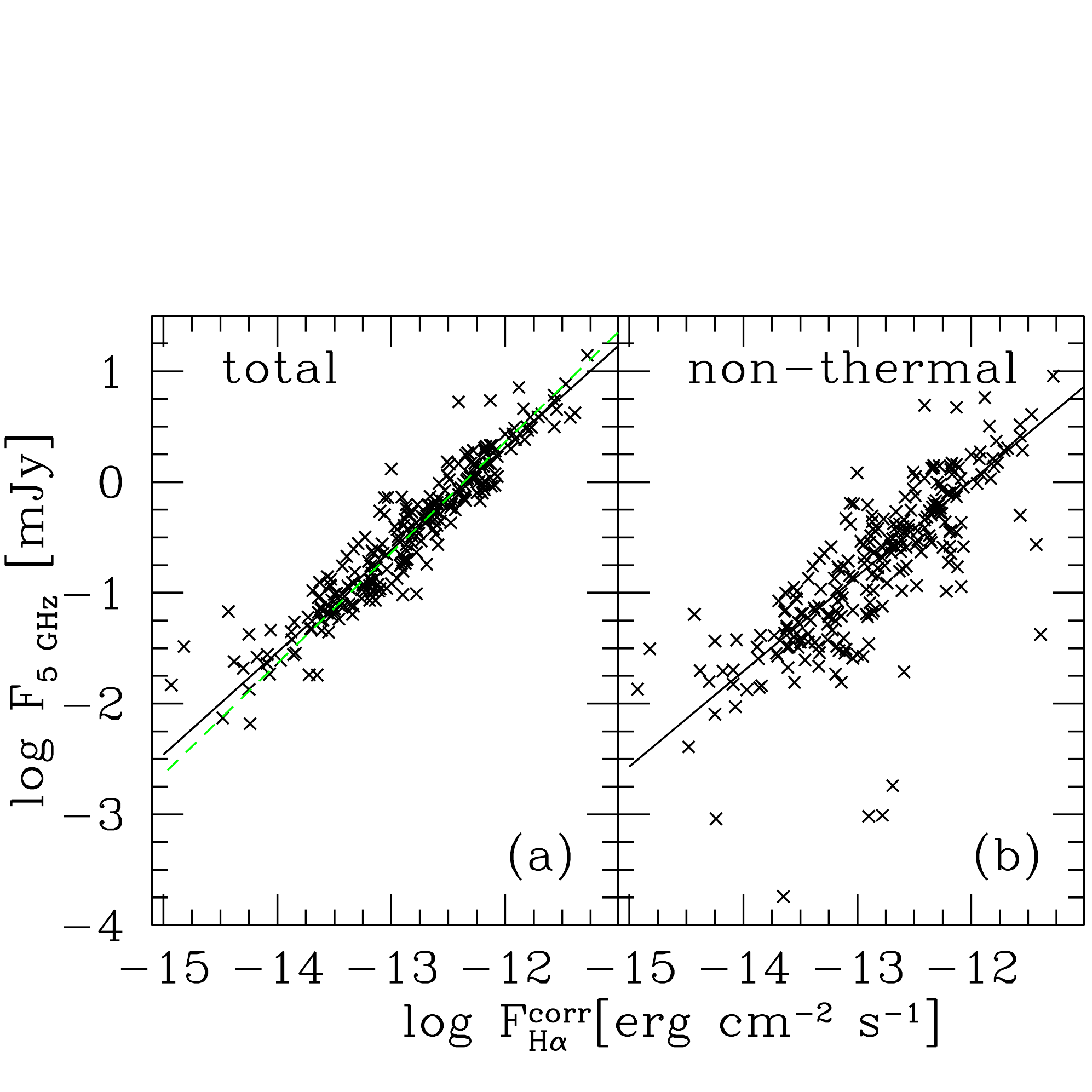}
 \caption{The total (left panel) and non-thermal radio flux  (right panel)
at 5~GHz for MIR selected star forming regions plotted as a function of the H$\alpha$ flux corrected
for extinction. The relation of the H$\alpha$ emission
 is tighter with the total radio flux rather than with the non-thermal radio continuum only. The fitted  
correlations shown are drawn by minimizing distances to the line and their   slopes are
0.87 and 0.93 for the total and non-thermal flux respectively.  Plotted quantities are relative to  circular apertures 
with R$_{ap}$=1.5 R$_s$. The green dashed line is the best fitted line with slope unity.
  }
\label{sfr}
\end{figure}

The distribution of the total radio flux at 5~GHz versus the   H$\alpha$ flux corrected for extinction,
 is shown in Figure~\ref{sfr} in a log-log plot.  The  straight line, a fit to the data minimizing distances,  has a slope 0.93  and  a very low dispersion. The dashed 
 green line is the fit to the data if we require a linear relation (a slope of one). The goodness of this fit implies that low mass star forming regions with 
 H$\alpha < 10^{-13}$~erg~cm$^{-2}$~s$^{-1}$  or  F$_{5GHz}<0.3$~mJy are equally dim in radio as in H$\alpha$ emission. 
The correlation of H$\alpha$ brightness with the non-thermal radio flux density shown by the straight line in 
 Figure~\ref{sfr}$(b)$  has a slope of  0.87 and is  less tight.  This suggests the use of total radio continuum,  which does not suffer extinction, 
 as indicator of the  SFR.  The following equation \citep{2012ARA&A..50..531K}  links the star formation rate with H$\alpha$ luminosities:
 
 \begin{equation}
{SFR_{H\alpha}\over M_\odot yr^{-1}}=5.37\ 10^{-42} {L_{H\alpha}\over erg~s^{-1}} = 4.53\ 10^{-4} {F_{H\alpha}\over  10^{-12} \hbox{ergs}~\hbox{cm}^{-2}~\hbox{s}^{-1}}
\end{equation}  

This is in general an upper limit  due to   ionizing photon leakage from star forming regions, a problem which  affects also
the thermal radio continuum. Using the linear relation (green dashed line) shown in Figure~\ref{sfr},   we can  write

\begin{equation}
{SFR_{5GHz}\over M_\odot yr^{-1}}= 1.98\ 10^{-4} {F_{5Ghz}\over  \hbox{mJy} }= 2.35\ 10^{-28} {L_{5GHz}\over erg~s^{-1}~Hz^{-1}}
\end{equation}  

We obtain the same expression when fitting the radio continuum flux density at 5~Ghz  versus H$\alpha$ emission in GMCs. We can then compare this expression,
which applies to discrete events of star formation at small scales, with the SFR  integrated over galaxies. Using the  FIR-radio continuum correlation  
to calibrate the star formation rate in galaxies as a whole \citep{2011ApJ...737...67M} this  reads:

\begin{equation}
{SFR^{large}_{1.4GHz}\over M_\odot yr^{-1}}= 0.64\  10^{-28} {L^{large}_{1.4GHz} \over erg~s^{-1}~Hz}
\end{equation}

Given the spectral indexes observed for the radio sources ($>-0.5$) the ratio between YSCs luminosities at 1.4 and at 5~GHz is $<2$.  Hence, for a given SFR,  equation (12) implies
a higher radio continuum luminosity than equation (11). That is to say that the ratio between the H$\alpha$ luminosity and radio continuum luminosity
is  higher in individual star forming regions than globally, for integrated quantities in spiral galaxies.  We recall that we reached a similar conclusion for the 
ratio between the MIR emission and radio continuum  in Section 3, with the exception of a few HII regions. On the other hand it is well known that  large scale radio emission 
in spiral galaxies has a strong non-thermal radio component \citep{1990ApJ...357...97C}:  magnetic fields are pervasive in the ISM and
cosmic rays can quickly diffuse away from where they are injected, supporting our finding.   
  
To investigate in more detail the link between the radio emission and the star formation rate, or  the ratio of thermal to non-thermal radio continuum 
flux density, and how these vary from  the small to  the large scale,  we now analyse the sample of GMCs. Cloud complexes  generally extend over about 100~pc,  
occupying a larger area than  star forming and HII regions,  except  for very massive YSCs  whose HII regions are very extended and have dispersed their native clouds.

\begin{figure}
\includegraphics[width=9. cm]{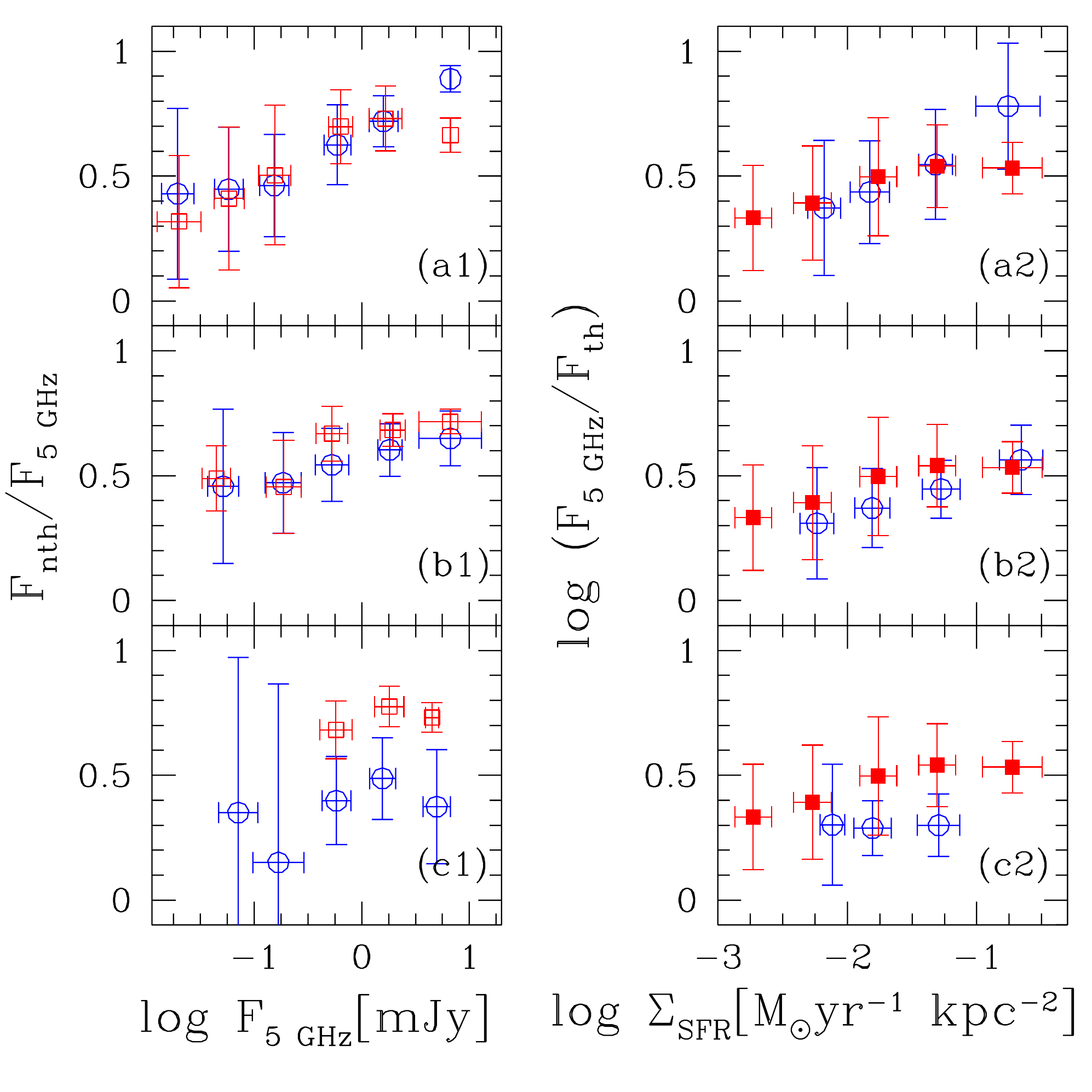}
 \caption{The left panels  show  the non-thermal to total radio continuum flux density ratio at 5~GHz  in radio sources associated to exposed YSCs (blue circles) using  apertures  with R$_{ap}$=1.5R$_s$
 for small $(a1)$, medium $(b1)$  and large $(c1)$ sources. In the same panels we show the non-thermal to total radio flux density ratio  for  GMCs hosting the YSCs (red squares), as a function
 of the radio emission of the YSCs. Not all YSCs have an associated GMC, especially  if they are at large galactocentric radii or spatially very extended.
  In the  right panels we plot the total-to-thermal radio flux ratio for the whole ensemble of GMCs hosting exposed YSCs (filled symbols), binned according to their  
  star formation rate per unit area.  Radio continuum has been used to estimate  the SFR density (red squares). 
   From top to bottom we compare the GMC ratios with that for the small, medium and large sources 
  respectively (open blue circles) binned according to their SFR per unit area. The area around the source  is circular  with a radius  1.5~R$_s$. 
  }
\label{sfrclo}
\end{figure}

In the left panels of Figure~\ref{sfrclo} we have added to the YSC data, already displayed in panels $(a1)$-to-$(c1)$ of Figure~\ref{size}, the mean GMC non-thermal to total
radio continuum ratios. These are computed for each bin of  YSC radio continuum flux density considering only molecular clouds associated with the YSCs
in that bin. They are shown as a function of the mean radio continuum flux density of the hosted YSCs in each bin.   In the middle and top panels, i.e. for small and medium size
YSCs, we can see that radio bright YSCs are hosted by GMCs with similar non-thermal to total radio continuum flux ratios. A few bright compact YSCs  have higher non-thermal 
to radio continuum flux ratios than their host GMCs (Fig.~\ref{sfrclo}$(a)$). This confirms our earlier finding, showed in
panels $(a2)$  of Figure~\ref{size}, that the high  non-thermal fractions in these sources decrease when considering a wider area around them.   
We cannot exclude that the localized excess of non-thermal fractions for a few bright compact radio sources is indeed not correct, but due to an underestimate of the free-free 
emission which has been considered optically thin \citep{1981A&A...102..424F,1999ApJ...527..154K,2001ApJ...559..864J}. 

 For extended radio sources,  Figure~\ref{sfrclo}$(c1)$  shows that although the associated GMCs follow the same trend with radio flux density
as the associated YSCs,  they have higher non-thermal fractions,  similar to clouds shown in  other panels and hosting more compact
sources.  In this case the HII region is comparable to or larger than the GMC and it is expanding while  dispersing the  parent cloud.  It is     
conceivable  that GMCs associated with some of these sources are leftovers or newly formed,  possibly triggered by  HII shell expansion.

In the right panels of Figure~\ref{sfrclo} we plot in red the total-to-thermal radio flux ratio for the whole ensemble of GMCs hosting exposed YSCs (filled red squares). 
Data has been  binned according to cloud  star formation rate per unit area and the same data  is displayed in all three panels for a direct comparison with the same ratio
measured for small $(a2)$, medium $(b2)$, and large $(c2)$  sources (open blue circles).  On the x-axis we show the SFR per unit area,
as traced by radio emission, computed inside cloud contours for GMCs, and  in a circular  area  with radius 1.5$R_s$ for YSCs. 
On the y-axis we display  the log of total-to-thermal  flux ratio for a direct comparison with  the model discussed by \citet{2016A&A...593A..77S}  (see their Figure~10).
For discrete star forming sites, such as those we are considering here, the Figure confirms  that the total-to-thermal radio flux ratio increases  with   
star formation density  for compact YSCs and clouds, but there is a flattening for clouds with $\Sigma_{SFR}> 0.03$~M$_\odot$~yr$^{-1}$~kpc$^{-2}$.  
The extended sources show a remarkably constant ratio of F$_{5GHz}$/F$_{th}$ of order 2 i.e. there is an equal share
between the thermal and non-thermal flux density.  The model of
 \citet{2016A&A...593A..77S} on a larger scale and for a continuous star formation predicts  an increase of F/F$_{th}$  with  $\Sigma_{SFR}$ faster than what we 
 observe for compact star forming regions or  GMCs.  The increasing trend is explained as due to turbulent magnetic field amplification by star formation. But, as the authors
 point out, discrete injection events cannot maintain a correlation between star formation rate and magnetic field strength. As a consequence,  and given also the
 higher frequency we are sampling, the slower increase of the total-to-thermal radio flux ratio with star formation in our  data is understandable.  
 This relation is lost if we trace the SFR density using  H$\alpha$ emission.
 This underlines the presence of fluctuations in the strength of radiative processes in individual YSCs, due to
 stochastic events  of massive star formation.

\subsection{Cosmic ray diffusion and the YSC contribution to non-thermal emission of M33}

At scales $<$1~kpc,   \citet{2013A&A...557A.129T} conclude that a turbulent
 magnetic field of order 8~$\mu$G dominates over the ordered component  in M33, and this tangled field implies that cosmic rays during their lifetime diffuse 
 over a relatively short pathlength, $<$400~pc.  The CRs are expected to be transported via streaming instabilities at the Alfven speed $V_A=B/\sqrt{4\pi\rho_{gas}} $.
 The average gas volume density in the giant cloud complexes we are considering  is of order 10$^{-23}$~gr~cm$^{-3}$ although individual molecular clouds in them,
 where stars form, have higher densities. For the duration of the star formation phase in GMCs, of
 order 10~Mys, the CRs can then diffuse over 70~pc, comparable to GMC sizes. 
 Bright and compact YSCs, which have not yet aged, are  producing massive  stars with discrete injection of CRs: these lag in the surrounding
cloud increasing the non-thermal fraction of the cloud radio emission.  
For extended HII regions, no longer surrounded by molecular gas, the gas density is lower and locally produced  CRs,  diffuse out quickly.  
As injections of CRs  stop local the non-thermal excess decays.

  What is the fraction of the global non-thermal emission of M33 that is in discrete sources such as  YSCs  or   SNRs ?
 \citet{2007A&A...472..785T} have measured the total  radio emission of M33 at 6.2~cm
 and find that the total flux is 1.28~Jy. The total H$\alpha$ flux of M33, corrected for extinction, is about 5 $10^{-10}$~erg~s$^{-1}$~cm$^{-2}$ \citep{2009A&A...493..453V},
 which implies that on average 1/3 of the emission of M33 at 6~GHz is thermal.  For a non-thermal spectral index of -0.8 \citep{2007A&A...475..133T},
 the total non-thermal emission at 5~GHz is about 1~Jy.  Supernovae in the radio catalogue of \citet{2019ApJS..241...37W} contribute about 8$\%$ to the total non-thermal flux,  
 at 5~GHz, while twice as much, about 16$\%$, is localized around YSCs hosting no known SNRs.  We conclude that about 1/4 of the total non-thermal emission 
 at 5~GHz in M33 is linked to discrete  current injection events of CRs  connected with massive stars evolution.
Diffusion of CRs into the ISM from ongoing and past generation of massive stars must account for the remaining fraction. 

The mean global star formation density in M33 is 0.0032~M$_\odot$~yr$^{-1}$~kpc$^{-2}$,  and the ratio of total to thermal flux is 5.7 at 1.4~GHz \citep{2007A&A...475..133T}. 
This estimate is in very good agreement  with  that predicted by  \citet{2016A&A...593A..77S} model for a continuous star formation law.

\section{Summary}

The formation of stars can be traced using UV stellar continuum, gas  recombination lines such as H$\alpha$ or radio thermal and non-thermal continuum, but also
 mid-infrared emission of hot dust grains which absorb radiation of the newly formed stars.
Star forming galaxies show a tight linear relation between their IR emission and radio continuum luminosities. 
High angular resolution surveys of the whole star forming disk of the closest galaxies   allow today to investigate such relations in individual molecular clouds 
as  the formation of  YSCs proceeds, and down to spatial scales below 50 parsecs.
In this paper we explore the  mid-infrared, radio continuum and H$\alpha$  emission for a large sample of star forming regions in M33 (526 YSCs)  and  for their
native molecular clouds. The YSCs candidates have been  selected for their mid-infrared  
 hot dust emission at 24~$\mu$m \citep{2011A&A...534A..96S}.  Recent radio surveys  of the nearby galaxy M33 have provided a detailed  view of the interstellar medium in 
 the radio continuum 
 and  a radio source catalogue in the M33 sky area \citep{2019ApJS..241...37W}.  Using this catalogue we  find  radio continuum 
source counterparts  at 1.4 and 5~GHz  for more than half of the MIR  sources. We use high sensitivity maps at at 6.3~GHz to measure successfully
radio emission for the most diffuse and weak ones without a catalogued counterpart.   

The radio continuum luminosity  of  a star forming region establishes a correlation with the 24$\mu$m luminosity of the associated hot dust component, 
whose slope is sublinear  in the log-log scale. 
The 24$\mu$ luminosity of YSCs fully embedded in the native molecular clouds follows a similar correlation with the radio continuum. We underline that
previous attempts to discover embedded star forming sources in M33 through radio continuum selected candidates   data have not been successful \citep{2006ApJS..162..329B}. 
Thanks to sensitive radio continuum maps at 6.3~GHz we have been able to confirm, for the first time in a nearby spiral galaxy, the embedded nature of many YSCs associated 
with molecular clouds with no optical or UV counterparts. Given  the variations of the spatial distribution of grains in the HII regions and of  the stellar continuum
fraction absorbed by dust, it is quite surprising to have found a relation that holds for the overall YSC population, from the embedded to the exposed phase, from the central region 
to the galaxy outskirts. 

The slow chemical enrichment of galaxies with cosmic time and the variety of processes that regulates life cycle and the temperature of dust grains implies  that 24~$\mu$m emission alone 
might not be sufficient to trace star formation.  A
combined tracer, IR with UV or hydrogen recombination lines, has become one of the most commonly used star formation rate estimator \citep{2009ApJ...703.1672K}. 
We have shown in this paper that the correlation between the radio continuum at 5~GHz and the 
combined tracer  24$\mu$m and H$\alpha$ emission (or extinction corrected H$\alpha$), holds over 4 orders of magnitude   at scales between  few tens  and $\sim$100~pcs. 
The relation is close to linear and extremely tight when data on YSCs refer only to the latest stage of evolution in the host molecular cloud (exposed phase).  
The correlation is   tighter  if the total radio continuum is considered rather than  the estimated non-thermal radio emission.
In examining  individual YSCs  of low luminosities, stochastic sampling of the IMF at its high mass end   increases the dispersion in the relation between 
hot dust emission and non thermal radio continuum due to stronger variations of  turbulent magnetic field amplification and CR production. 

We find a significant fraction of the 5~GHz emission being thermal but    compact radio bright YSCs have more than half of their radio emission due to non thermal processes.
 The unambiguous non-thermal emission in star forming regions where the H$\alpha$ flux is greater than 10$^{-13}$~erg~s$^{-1}$~cm$^{-2}$, even in the absence of a supernova remnant,
 implies ubiquitous local production of cosmic rays when the YSC has at least  an O7-type star. Indeed theoretical computations predicts that these O-type stars
 are able to produce a wind and therefore a shock.   About 1/4 of the total non-thermal emission of M33 
 is linked to discrete  current injection events of CRs  connected with massive stars in YSCs or SNRs. 
Diffusion of CRs into the ISM from ongoing and past generation of massive stars must account for the remaining fraction.
 
 More extended radio sources, likely associated with  evolved HII regions, do not
 show a correlation between non-thermal fraction and radio luminosity, and have on average lower non-thermal fractions than  more compact ones. 
 We interpret this trend as due to fast diffusion of CRs  as the HII shell expands well beyond the native cloud,  with CR injections  becoming more seldom at a later stage of 
 the cluster evolution. CR diffusion also explain the lower ratios
  between the radio continuum  at 5~GHz and  the 24$\mu$m  or  H$\alpha$ emission in individual star forming regions of M33  compared to those measured
for  galaxies as a whole in the local universe.  

 The radio continuum emission from giant molecular complexes  follows a similar correlation  with the extinction corrected H$\alpha$  
as individual star forming regions.  As determined by \citet{2017A&A...601A.146C}  the duration of the life cycle for M33 molecular clouds   is about 14~Gyrs with  the mean molecular cloud mass that increases only by 
a factor 3 going from the initial inactive phase to the embedded one, to the exposed phase prior to cloud dispersal.  The radio emission of the whole GMC has a   stronger increase as star formation progresses: as  the YSC is formed and it starts breaking through the cloud   the non-thermal component in the cloud is ubiquitous. 
Molecular clouds with only embedded sources,  or molecular clouds hosting very small YSCs, as well as inactive clouds,
have much lower radio continuum luminosities on average:   many of them are still undetected in the published surveys.  

The  non thermal-- radio continuum correlation observed  for compact YSCs and GMCs with exposed star formation at all galactocentric radii underlines the role of the star-formation induced 
 turbulent magnetic field as the most efficient source of the non thermal radio emission. Tangled magnetic field in  turbulent molecular complexes  can prevent   
 fast diffusion of cosmic ray electrons to larger scales as suggested by  \citet{2013A&A...557A.129T}.   
This can be better investigated in the future using  adequate deeper radio survey for tracing the  diffuse radio emission.
 We have calibrated the radio continuum
emission as star formation tracer  at scales of individual star forming regions and GMCs in M33.  A shallow increase of the average total to thermal radio flux ratio with  the
star formation density is  measured for compact YSCs as well as for  giant molecular complexes. This is in agreement with theoretical models of turbulent magnetic field amplification by star formation, 
although fluctuations in CR injections  during localized  star formation events  require more specific  theoretical modeling. This will shed light  on the link between small and large scale radio 
emission in galaxies.

\begin{acknowledgements}
EC acknowledges the support from grant PRIN MIUR 2017 -$ 20173ML3WW_001$  and from the INAF PRIN-SKA 2017 program 1.05.01.88.04.
\end{acknowledgements}


%
\end{document}